\documentclass[sn-mathphys]{sn-jnl}

\jyear{2022}%

\theoremstyle{thmstyleone}%
\theoremstyle{thmstyletwo}%

\theoremstyle{thmstylethree}%

\begin{document}

\title[Article Title]{In-orbit performance of HE onboard  \textit{Insight}-HXMT in the first 5 years}



\author[1]{\fnm{Xu-Fang} \sur{Li}}

\author*[1]{\fnm{Cong-Zhan} \sur{Liu}}\email{liucz@ihep.ac.cn}
\author[1]{\fnm{Yi-Fei} \sur{Zhang}}
\author[1]{\fnm{Xiao-Bo} \sur{Li}}
\author[1]{\fnm{Zheng-Wei} \sur{Li}}
\author[1]{\fnm{Xue-Feng} \sur{Lu}}
\author[1]{\fnm{Zhi} \sur{Chang}}
\author[1]{\fnm{Ming-Yu} \sur{Ge}}
\author[1]{\fnm{Juan} \sur{Zhang}}
\author[1]{\fnm{Yu-Peng} \sur{Xu}}
\author[1]{\fnm{Fang-Jun} \sur{Lu}}
\author[1]{\fnm{Li-Ming} \sur{Song}}
\author[1]{\fnm{Shuang-Nan} \sur{Zhang}}

\affil*[1]{\orgdiv{Key Laboratory of Particle Astrophysics}, \orgname{Institute of High Energy Physics, Chinese Academy of Science}, \orgaddress{\street{19B Yuquan Road, Shijingshan District}, \city{Beijing}, \postcode{100049},  \country{China}}}






 \abstract
 {\textbf{Purpose:} The High-Energy X-ray telescope (HE), one of the three main payloads of the \textit{Insight}-HXMT mission, is composed of eighteen NaI(Tl)/CsI(Na) phoswich detectors, where NaI(Tl) serves as the primary detector covering 20--250\,keV, and CsI(Na) is used as an active shield detector to suppress the background of NaI(Tl) and also serves as an all-sky gamma-ray burst monitor covering 0.2--3\,MeV. In this paper, we review the in-orbit performance of HE in the first 5 years since \textit{Insight}-HXMT was launched on June 15, 2017. 
 \textbf{Methods:} The major performances we concern include the gain and energy resolution of NaI(Tl) and CsI(Na) detectors, the performance of pulse-shape-discriminator (PSD) and system dead-time. In this work, we investigate these performances mainly using the data of blank-sky observations and the data when the telescope in earth occultation. 
 \textbf{Results:} The overall performance of HE/NaI(Tl) is very stable in the first 5 years, whereas the gain of HE/CsI(Na) shows a continuously increasing trend and should be calibrated regularly. %
 \textbf{Conclusion:} In general, HE is still in good health and well-calibrated status after five-year's operation. The in-orbit performance of HE has no significant deviation from expectation. HE is expected to be in operation healthily for another several years of extended mission life.}




\maketitle

\section{Introduction}\label{sec1}

The Hard X-ray Modulation Telescope (dubbed \textit{Insight}-HXMT after launch), China's first X-ray astronomy satellite, was launched on June 15, 2017\cite{Zhang2020, Liu2020}. There are three main science instruments onboard \textit{Insight}-HXMT, covering the energy range from 0.7\,keV to 3000\,keV. The High Energy X-ray telescope (HE), the biggest one among the three instruments, consists of  eighteen NaI(Tl)/CsI(Na) phoswich scintillation detectors (HEDs)\cite{Liu2020}. Each detector has a collimator with an automatic gain control detector (AGC) inside. The whole instrument is shielded by six anti-coincidence detectors (HVT) on the top and twelve lateral anti-coincidence detectors. The fluorescence photons produced in the two types of crystals of HED are detected by a shared photomultiplier tube (PMT). Pulse height and width of scintillation signals are analyzed by \textit{Pulse Height Analyzer}(PHA) and \textit{Pulse Shape Analyzer}(PSA) respectively. Thanks to the pulse shape discrimination (PSD) technology, the NaI events and CsI events can be distinguished easily. For HED phoswich, the NaI(Tl) acts as the main science detector covering an energy band  from 20--250\,keV, and the CsI(Na) detector acts as an active shielding detector to reject Compton events which deposit energies in both CsI(Na) and NaI(Tl) simultaneously, and to reject background events incident from the backside within ~2 $\pi$ solid angle. However, the CsI(Na) detector can also be used as an all-sky Gamma-Ray Burst (GRB) monitor covering the energy range from 40--800\,keV in the normal mode and 0.2--3\,MeV in the low gain mode\cite{Zhang2020, Liu2020, Song2022}. 

Performances of HE have been derived from a combination of on-ground calibration results and dedicated calibration observations during its in-orbit operation. Detailed descriptions of its design, on-ground calibrations, and in-orbit calibrations were previously reported\cite{Liu2020,Li2019,Li2020,Luo2020}. This work will focus on the evolution of the performance of the HE phoswich detectors during the first 5 years in orbit. 

\section{Operation modes}\label{sec2}
To meet the requirements of the scientific observations, HE was designed to have three operation modes:

(1) \textit{NOrmal Mode} (NOM): In this mode, with the help of auto-gain control system, the gain of a NaI(Tl) detector is continuously monitored and the negative \textit{high-voltage} (HV) of the detector's PMT is automatically adjusted, so that in NaI(Tl) detector the full-energy peak of the 59.5 keV photons (emitted from the radioactive source $^{241}$Am in AGC ) can be fixed at PHA channel 50. In this operation mode, HE can simultaneously observe sources within the Field of View (FoV) restricted by collimators in 20--250\,keV with NaI(Tl) and monitor almost the entire sky in 40--800\,keV with CsI(Na). 

(2) \textit{Low Gain Mode} (LGM): In this mode, the HE auto-gain control system is turned off and the HVs of the HED PMTs are reduced to make the energy range of CsI(Na) cover about 0.2--3\,MeV. This mode is usually scheduled when the \textit{Insight}-HXMT runs into the earth occultation where the observed sources or scanned regions are invisible and the normal observations have to be interrupted. 

(3) \textit{South Atlantic Anomaly} (SAA) mode: In this mode, the HVs of HEDs are switched off to protect the detectors from the damage of high-flux charged particles. The instrument will not record any scientific event. 

\section{Performance evolution in orbit}\label{sec3}

\subsection{High voltage monitoring}\label{subsec3-1}
\begin{figure}[H]%
\centering
\includegraphics[width=0.32\textwidth]{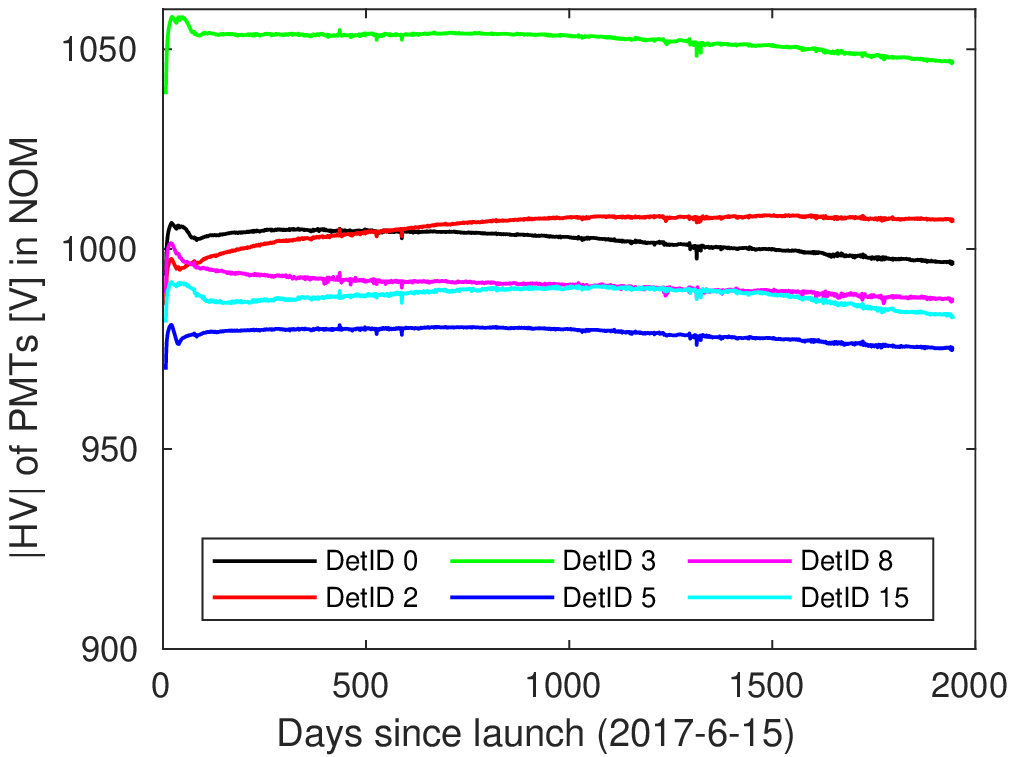}
\includegraphics[width=0.32\textwidth]{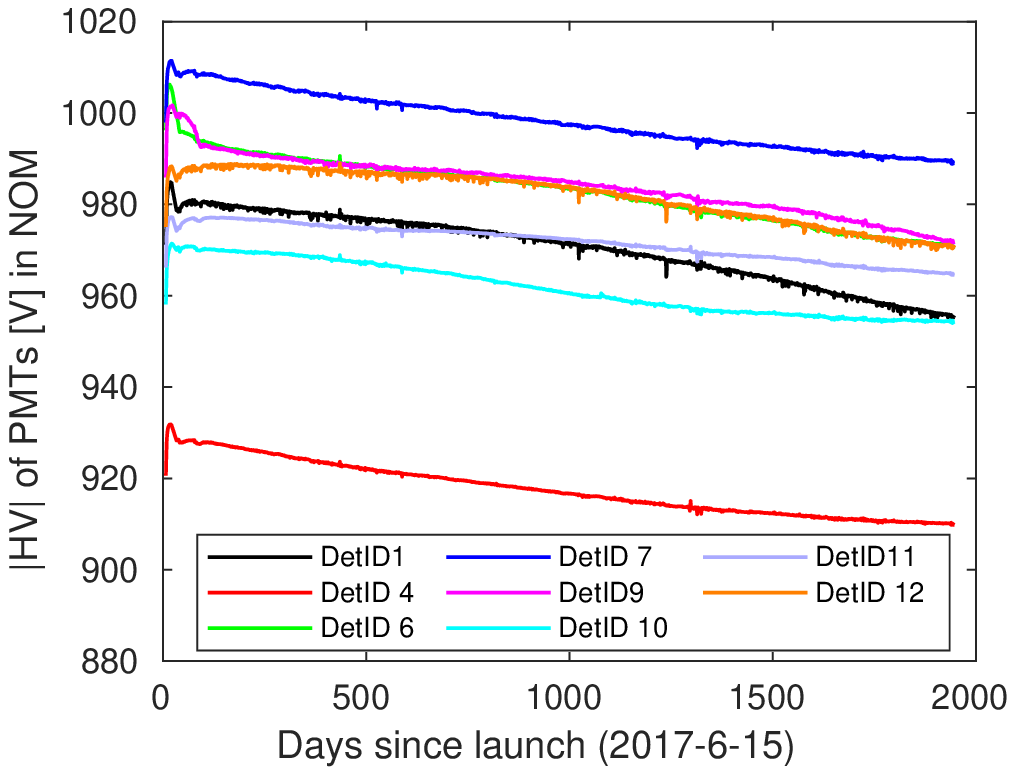}
\includegraphics[width=0.32\textwidth]{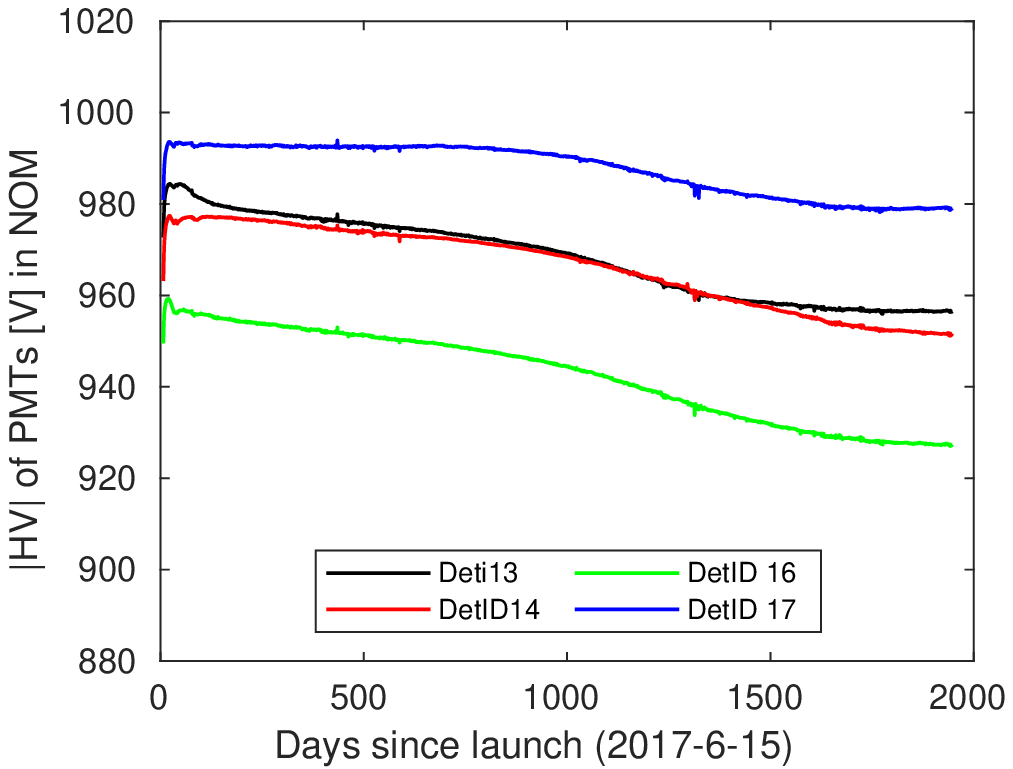}
\caption{Absolute values of HED HVs as a function of time in NOM in the first five years. The HVs are automatically adjusted by AGC system.}\label{fig1_HVinNOM}
\end{figure}

\begin{figure}[H]%
\centering
\includegraphics[width=0.45\textwidth]{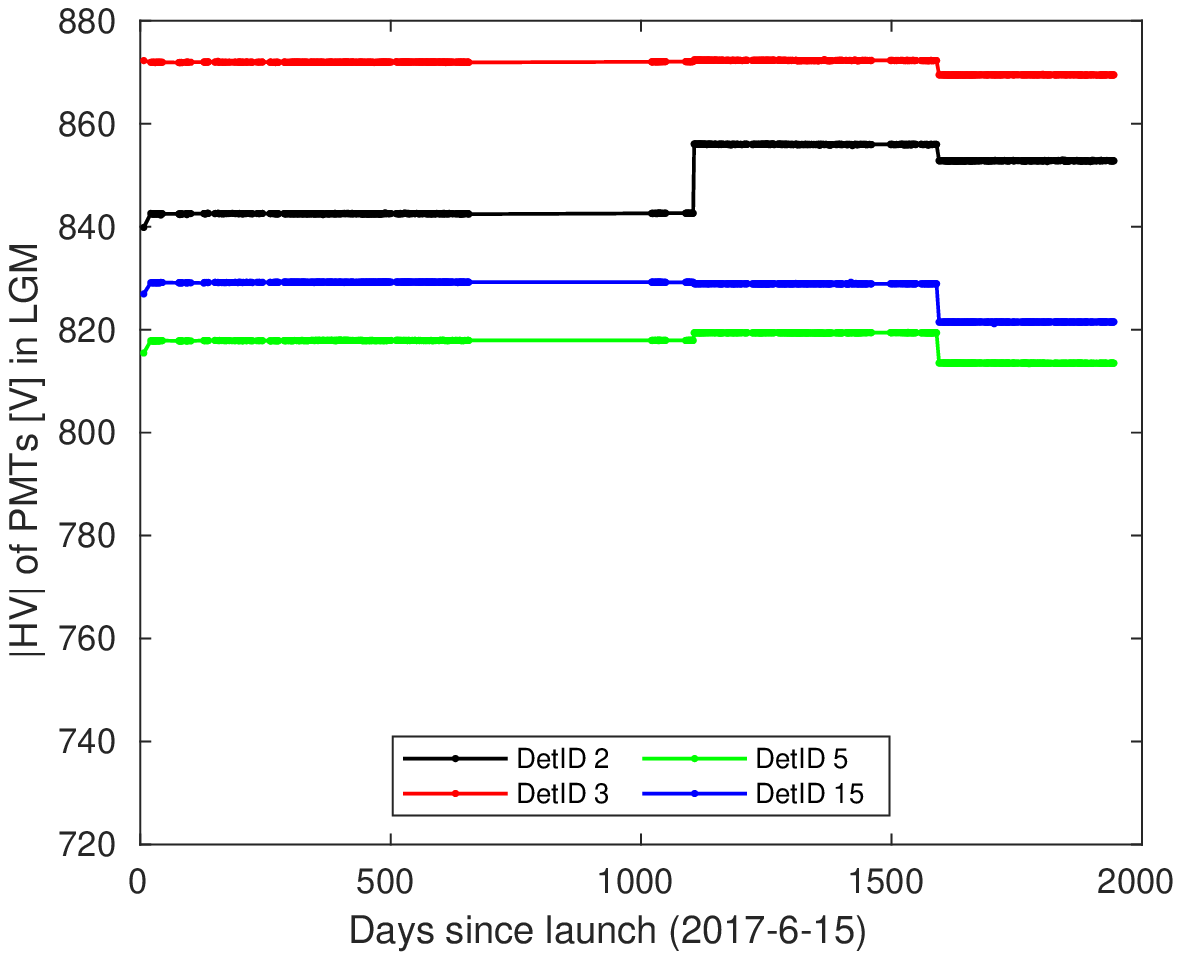}
\includegraphics[width=0.45\textwidth]{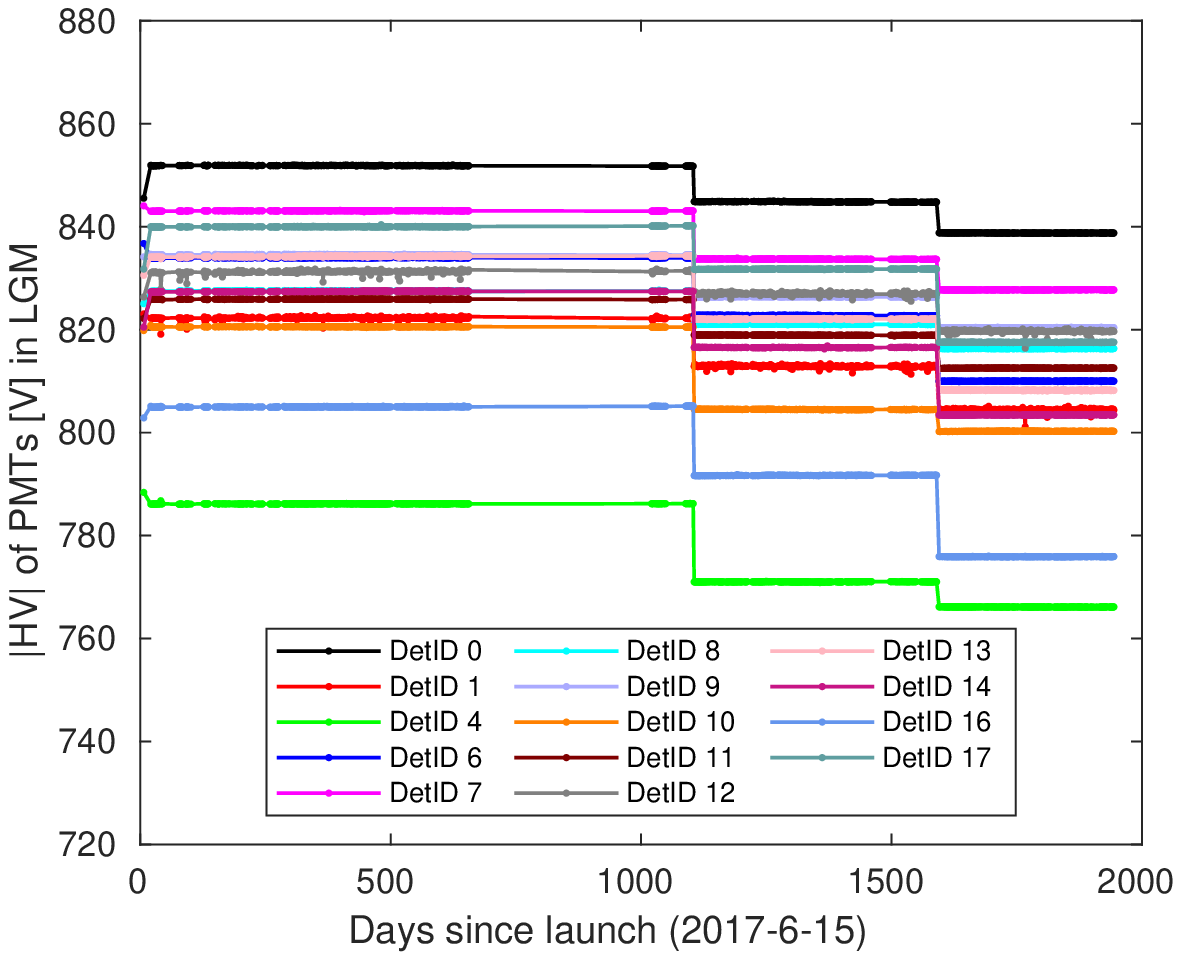}
\caption{Absolute values of HED HVs as a function of time in LGM in the first five years.The voltages were adjusted by tele-commands in July 2017, Jun. 2020 and Oct. 2021, respectively. }\label{fig2_HVinLGM}
\end{figure}

The HED was designed to work with negative HV (for convenience, we only use the absolute value of HV hereafter). Figure \ref{fig1_HVinNOM} and Figure \ref{fig2_HVinLGM} show the daily average absolute values of the PMT HVs in NOM and LGM over time, respectively. The HVs changed obviously during the first three months because the telescope was in commissioning phase. About three months later, the HED voltages tended to change slowly. In the middle and right panels of Figure \ref{fig1_HVinNOM}, we can see that most of HED HVs declined in the last five years with different slopes. In the left panel of Figure \ref{fig1_HVinNOM}, HVs of DetID 0,3,5 and 15 are certainly stable, but that of DetID 2 is peculiar with a slow increase. In LGM, the HVs can not be adjusted automatically because the AGC system is turned off. According to the calibration results, the HVs in LGM were updated in July 2017, Jun. 2020, and Oct. 2021, respectively, resulting in the three discontinuities in Figure \ref{fig2_HVinLGM}. Most of HED HVs in LGM decreased but DetID 2. From the left panel of Figure \ref{fig2_HVinLGM}, we can clearly see that HV of DetID 2 has been progressively increasing, which is coincident with the behavior in NOM.

In general, the gain of each HED detector system is determined by the fluorescence yielding and collection efficiency, the HV applied to the PMT and the gain of the electronics chain. In HE design, the gain of each HED detector system is fixed by the corresponding AGC system. And the gain of electronics chain is relatively stable because there is an active temperature control for each HED to keep the detector temperature not changed too much. So we suppose that the HV adjustment is a compensation for the change of fluorescence yielding and collection efficiency. Before launch we found the HEDs slightly deliquesced to some extent. The vacuum environment in orbit is beneficial to the HED scintillators recovering from deliquescing. Then the fluorescence yielding and collection efficiency improved slowly, leading to the gain increasing and the HV decreasing at the same time. But the behavior of DetID 2 can not be understood very well, which indicates the nature is still complicated.

\subsection{Performance evolution of gain and energy resolution}\label{subsec3-2}
For regular energy response calibration, we hope the background and source is stable and observation frequency is suitable. So the data of blank-sky observations are used to calibrate the energy responses of NaI(Tl) and CsI(Na) detectors in NOM, and the data of earth occultation are used to calibrate CsI(Na) in LGM. We select events with pulse width between PSA channels 54 to 70 as NaI events, and events with pulse width between PSA channel 80 to 120 as CsI events in NOM.  In LGM, we only select CsI events with pulse width between PSA channel 78 to 120. Those events with veto tags by HVTs are rejected. 

\subsubsection{NaI(Tl) performance}\label{subsec3-2-1}
Plenty of experimental studies on NaI(Tl) detector over many years have shown that the NaI(Tl) crystal's light output is not strictly proportional to the deposited energy of X-ray and electron\cite{NLRofNaI1963,Wayne1998,Gardner1999,Payne2011}. The non-proportional response (NPR) model of each HE/NaI(Tl) detector to electrons was built by simulations based on on-ground calibration campaign\cite{Li2019,Li2020}. According to the NPR model, the energy response of the NaI(Tl) to X-ray was determined with the help of Monte Carlo simulations. In orbit, four iodine activation lines (31, 56, 67, and 191\,keV)\cite{zhangjuan2020} are used to modify the on-ground calibrated Channel-Energy relation and energy resolution model. Then the NPR model and response matrix files of each HE/NaI(Tl) detector can be produced. The in-orbit calibration progress and results of the HE/NaI(Tl) were described in detail in\cite{Li2020} .

\begin{figure}[H]%
\centering
\includegraphics[width=1.0\textwidth]{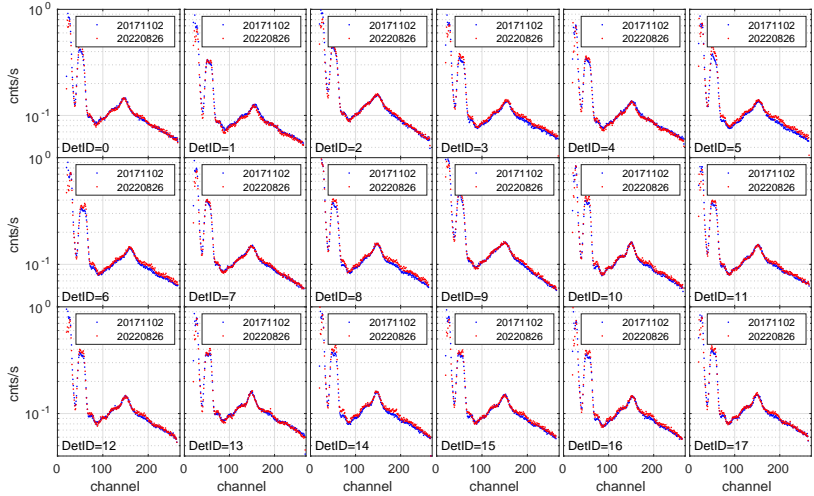}
\caption{Background PHA spectra of HE/NaI(Tl) detectors obtained from two blank-sky observations. In each panel, one curve is for 4.5 months after launch (blue) and another one is for 5 years after launch (red).}\label{fig3_NaIPHAcmp}
\end{figure}

Regular blank-sky observations and the $^{241}$Am radioactive sources in AGCs are used to monitor the long-term gain evolution of HE/NaI(Tl) detectors. Figure \ref{fig3_NaIPHAcmp} shows the energy spectra of eighteen HE/NaI(Tl) detectors of two blank sky observations. In each panel, the blue curve is the energy spectrum of 4.5 months after launch and the red curve is that of 5 years after launch. The two spectra are almost identical except the intensity of the bump around $\thicksim$67\,keV. The bump intensity increased with time, which is attributed to the in-orbit cumulative particle-induced activation of iodine isotopes in NaI(Tl) and CsI(Na) detectors with long decay periods. In the first year, the relative abundance of $\thicksim$67\,keV line increased quickly and eventually reached an equilibrium between activation and decay\cite{zhangjuan2020}. 

\begin{figure}[H]%
\centering
\includegraphics[width=0.45\textwidth]{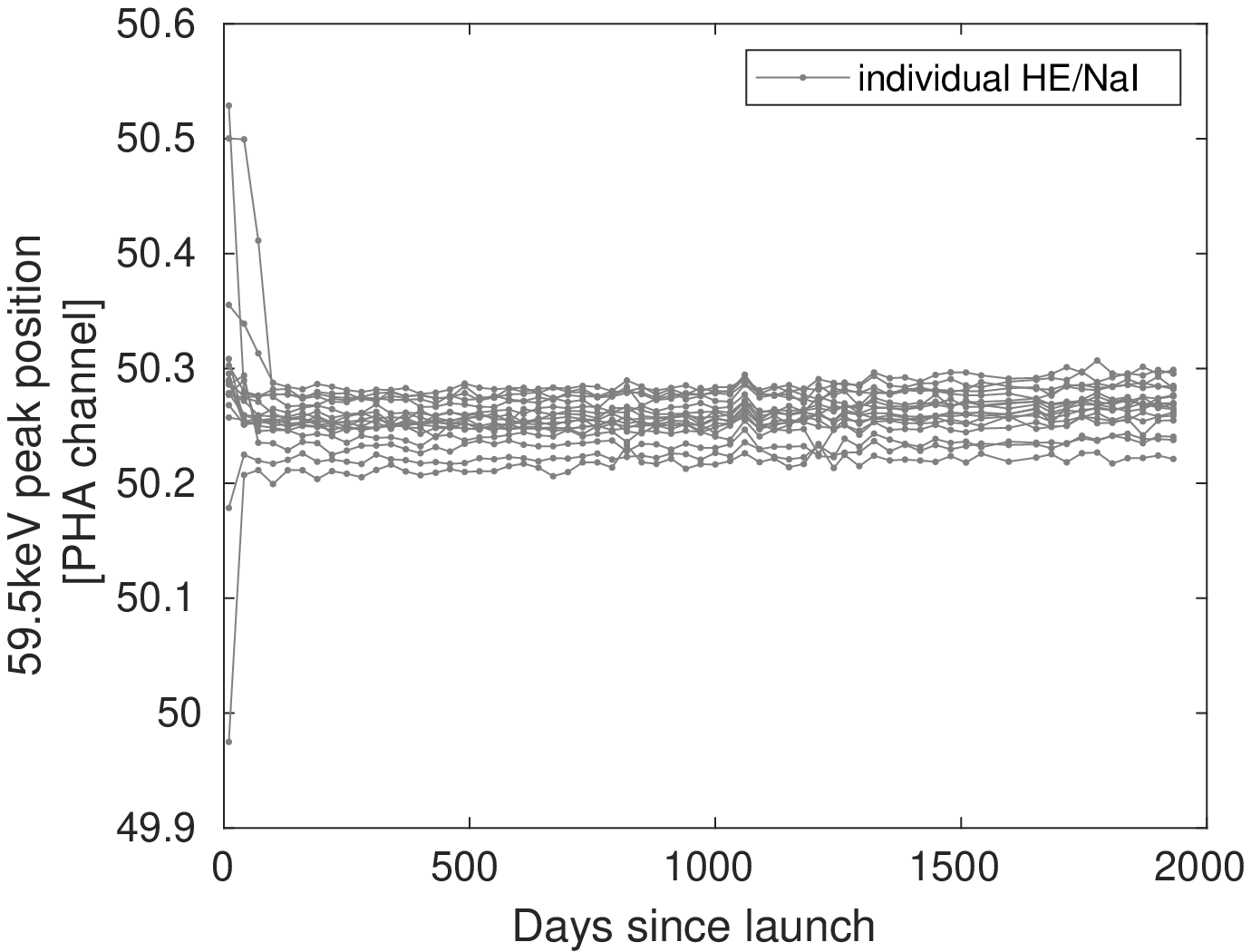}
\includegraphics[width=0.45\textwidth]{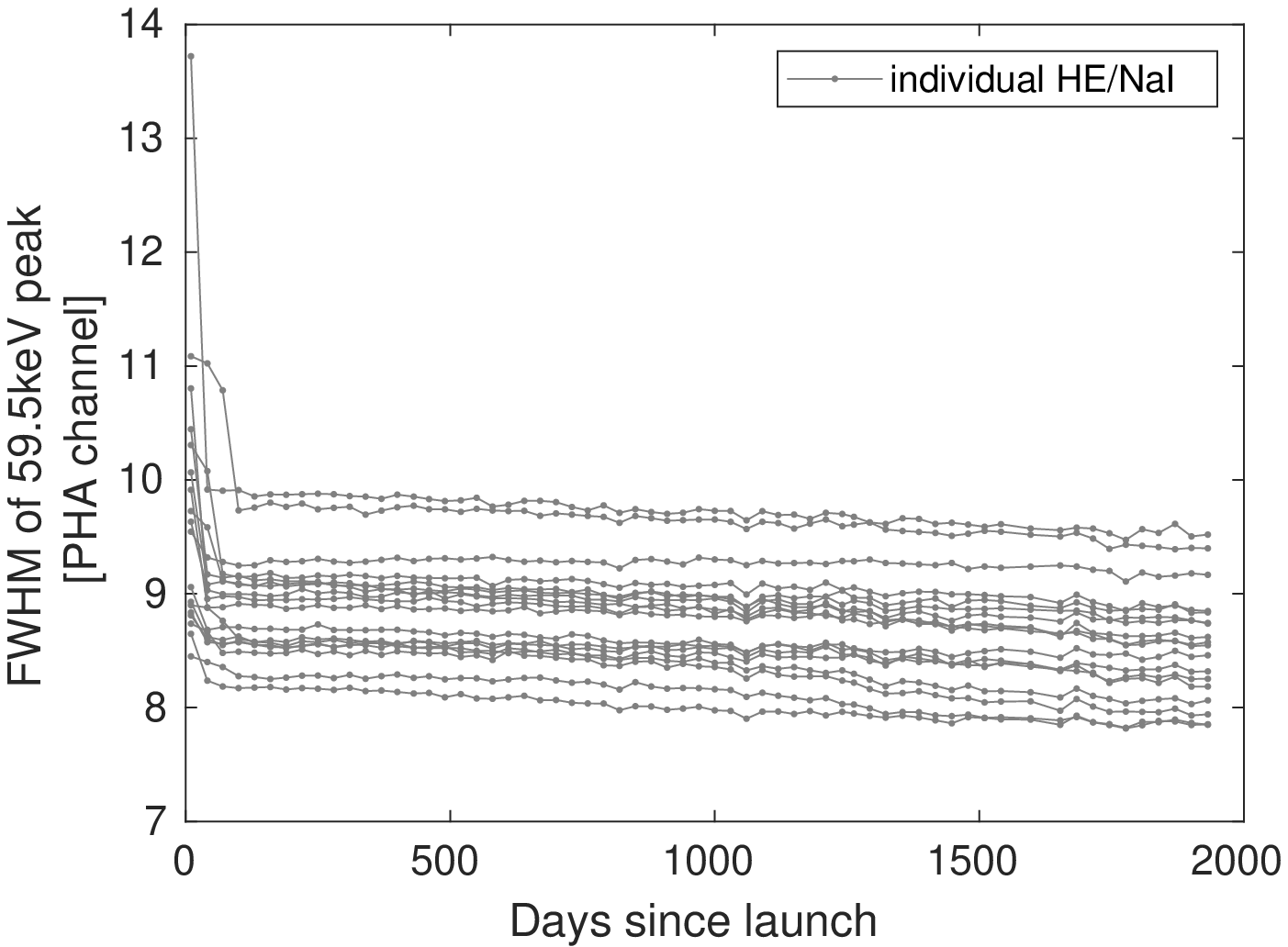}
\caption{The 59.5\,keV peak positions (left) and resolutions in FWHM (right) as a function of time for 18 HE/NaI(Tl) detectors.(a) Evolutions of the 59.5\,keV peak position for eighteen HE/NaI(Tl) detectors over time. (b) Evolutions of the energy resolution of 59.5\,keV peak for eighteen HE/NaI detectors over time.}\label{fig4_595peakinfo}
\end{figure}

\begin{figure}[H]%
\centering
\includegraphics[width=0.7\textwidth]{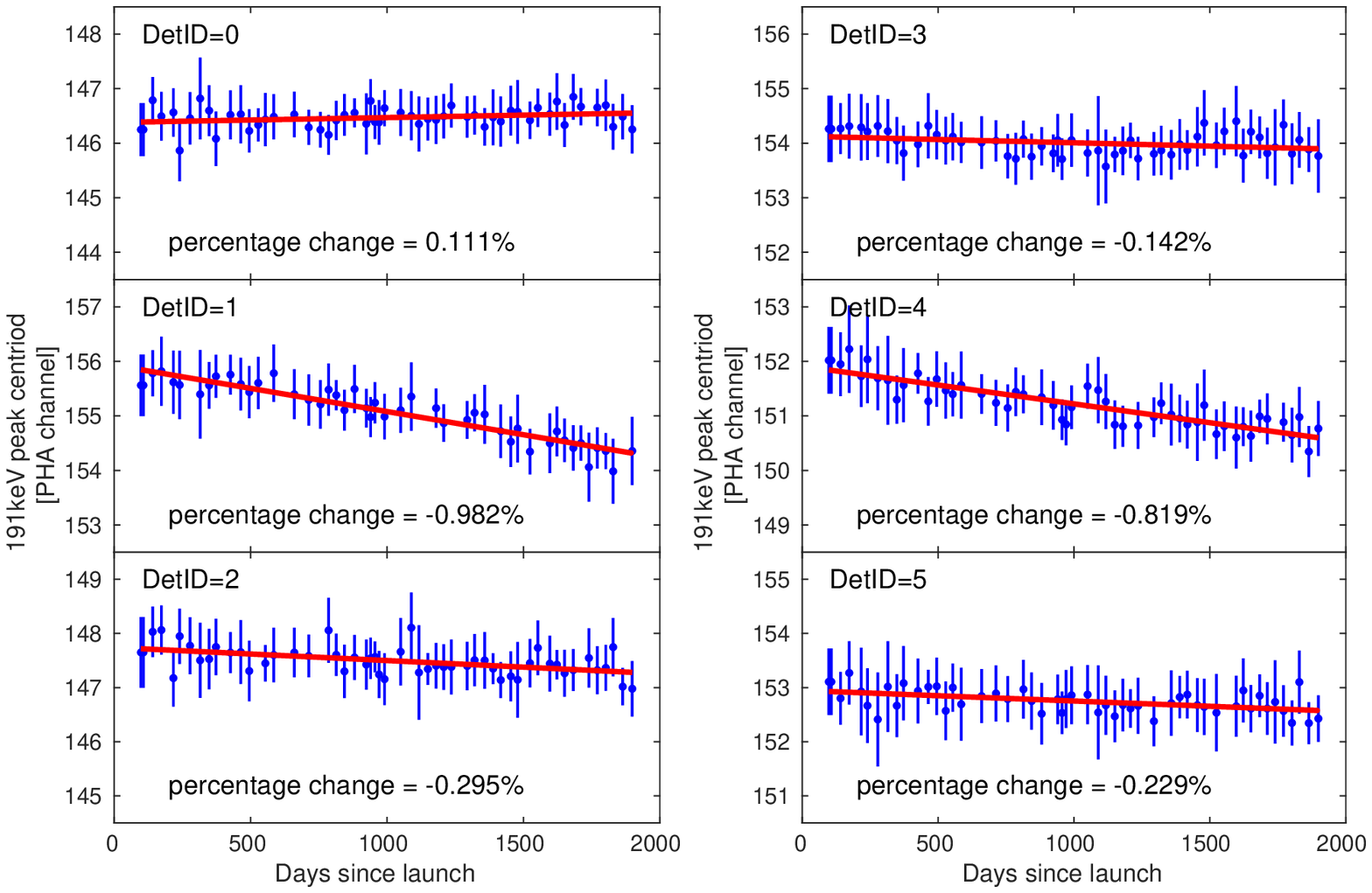}
\includegraphics[width=0.7\textwidth]{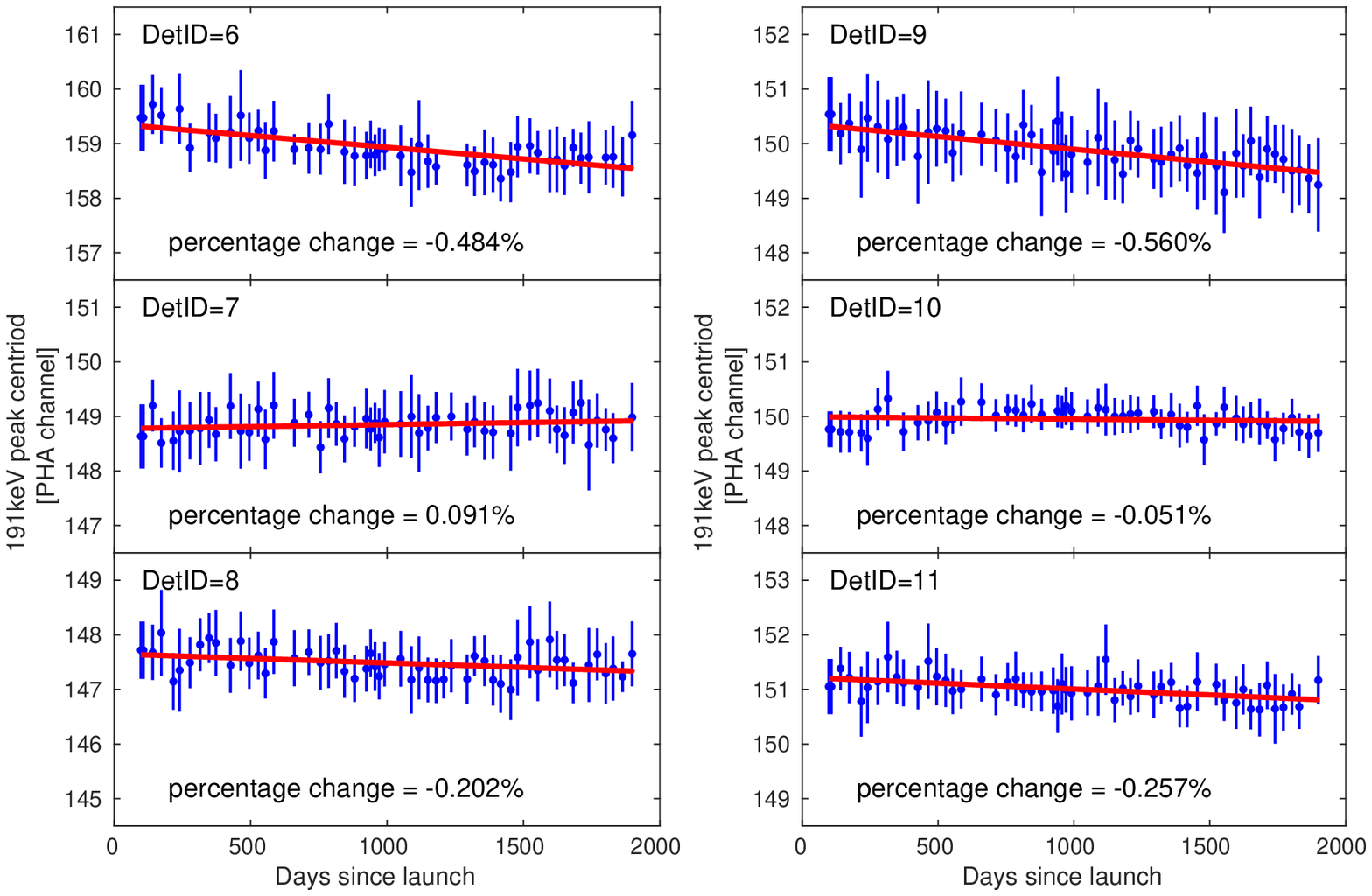}
\includegraphics[width=0.7\textwidth]{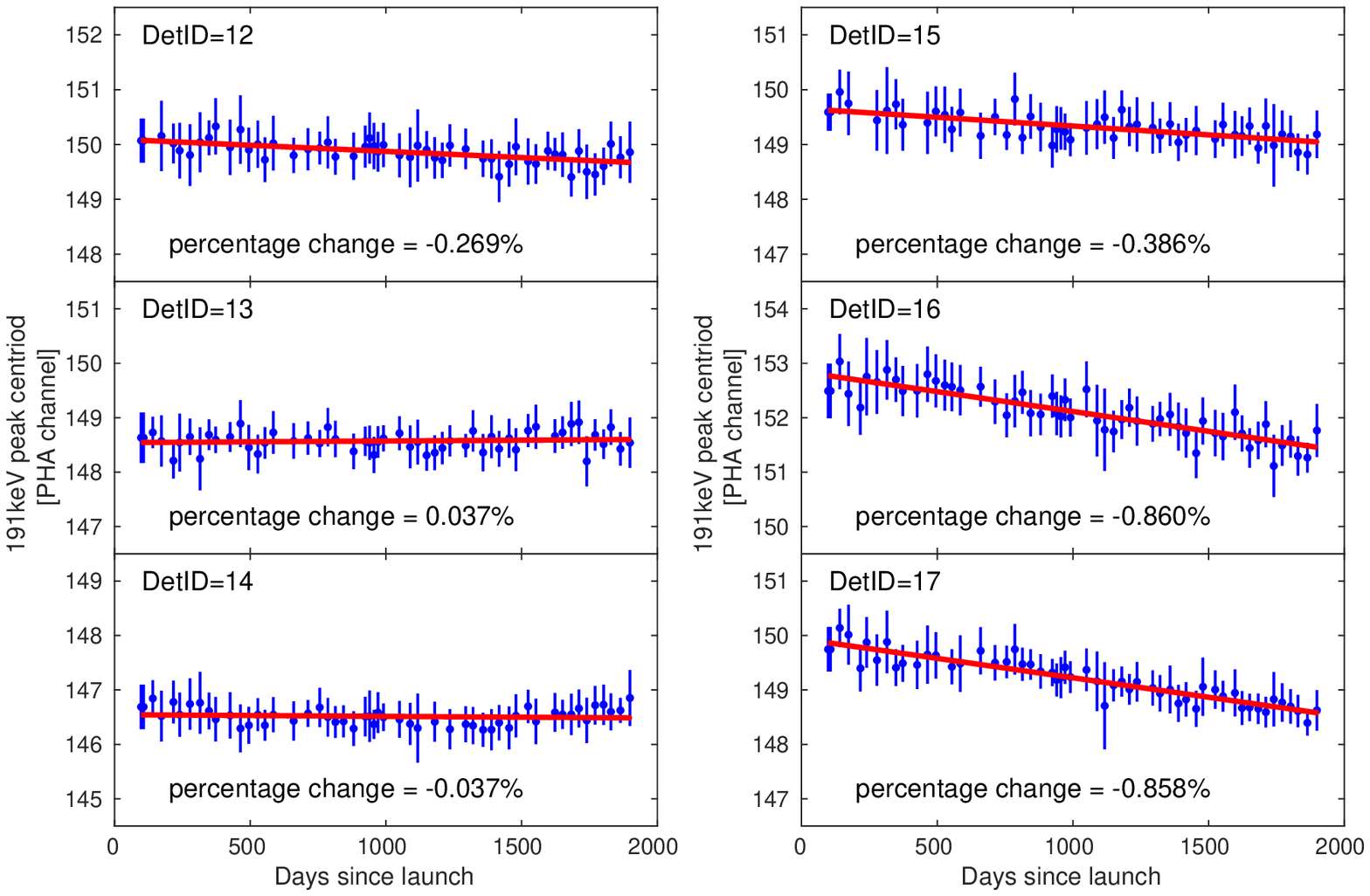}
\caption{Peak centroid of 191\,keV line  for 18 HE/NaI(Tl) detectors as a function of time. Compared with the values just after launch, the percentage change of the peak positions from Sep.29, 2017 to Jun. 26, 2022 are within 1$\%$.}\label{fig5}
\end{figure}

\begin{figure}[H]%
\centering
\includegraphics[width=0.7\textwidth]{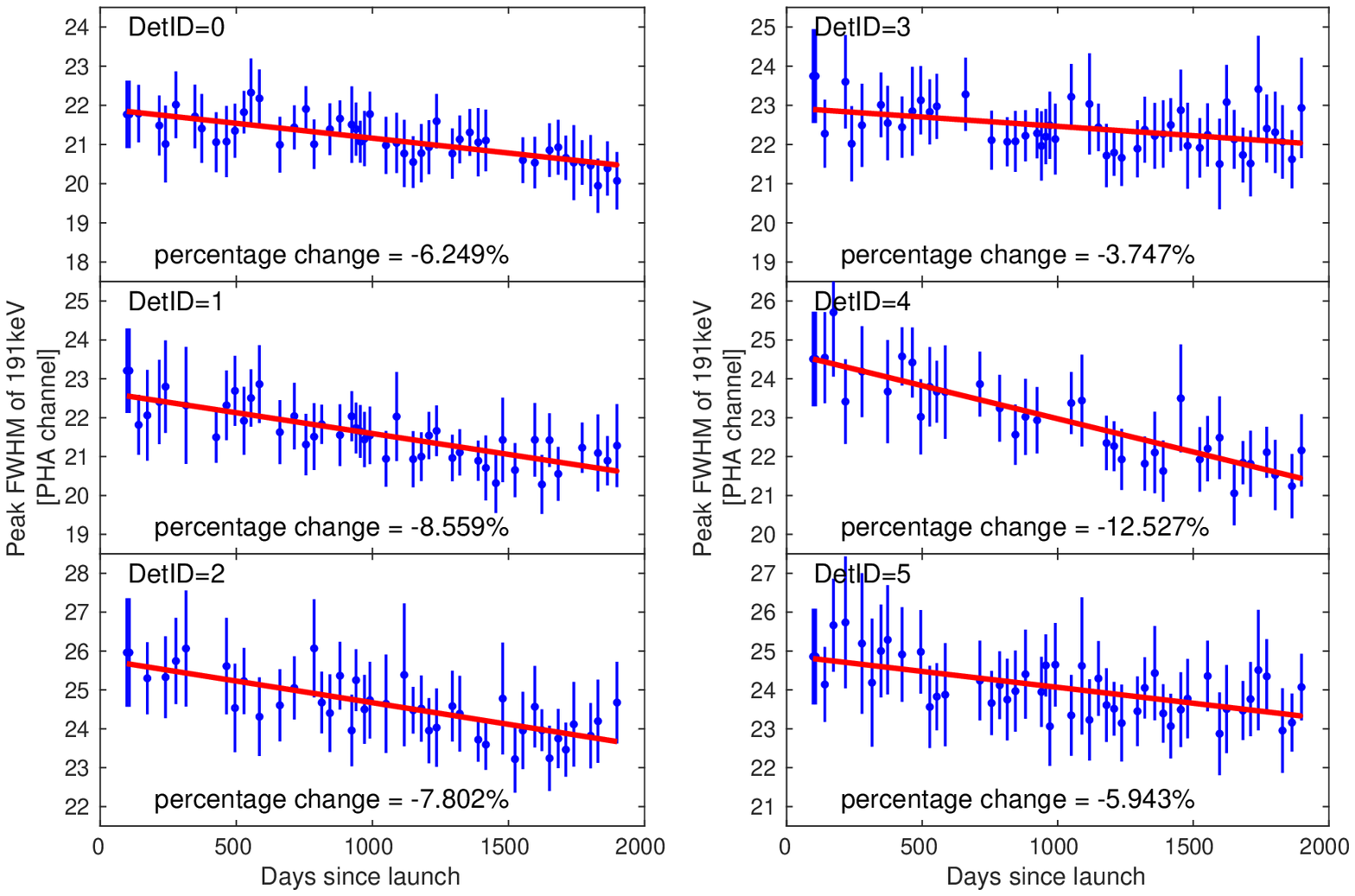}
\includegraphics[width=0.7\textwidth]{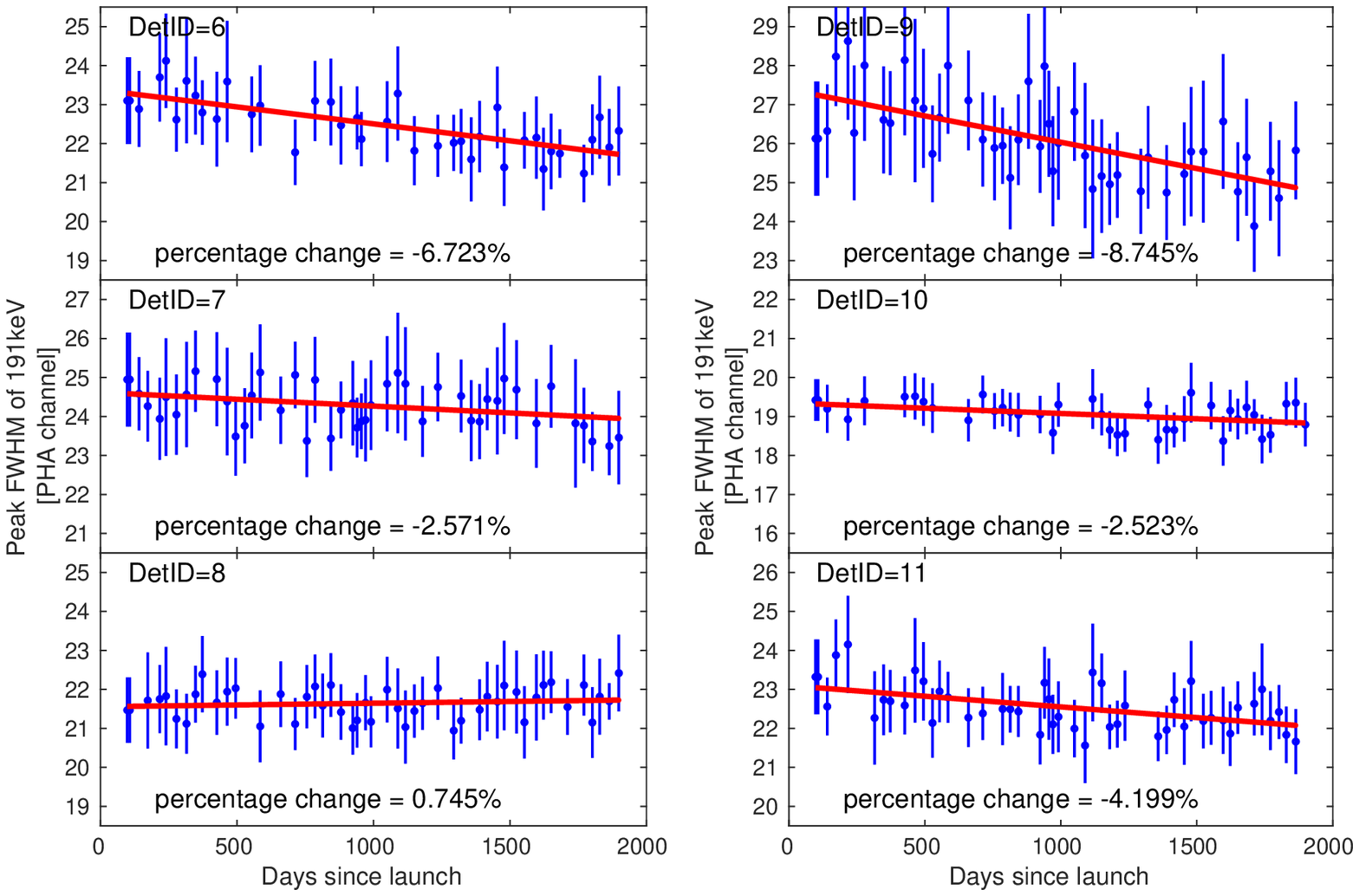}
\includegraphics[width=0.7\textwidth]{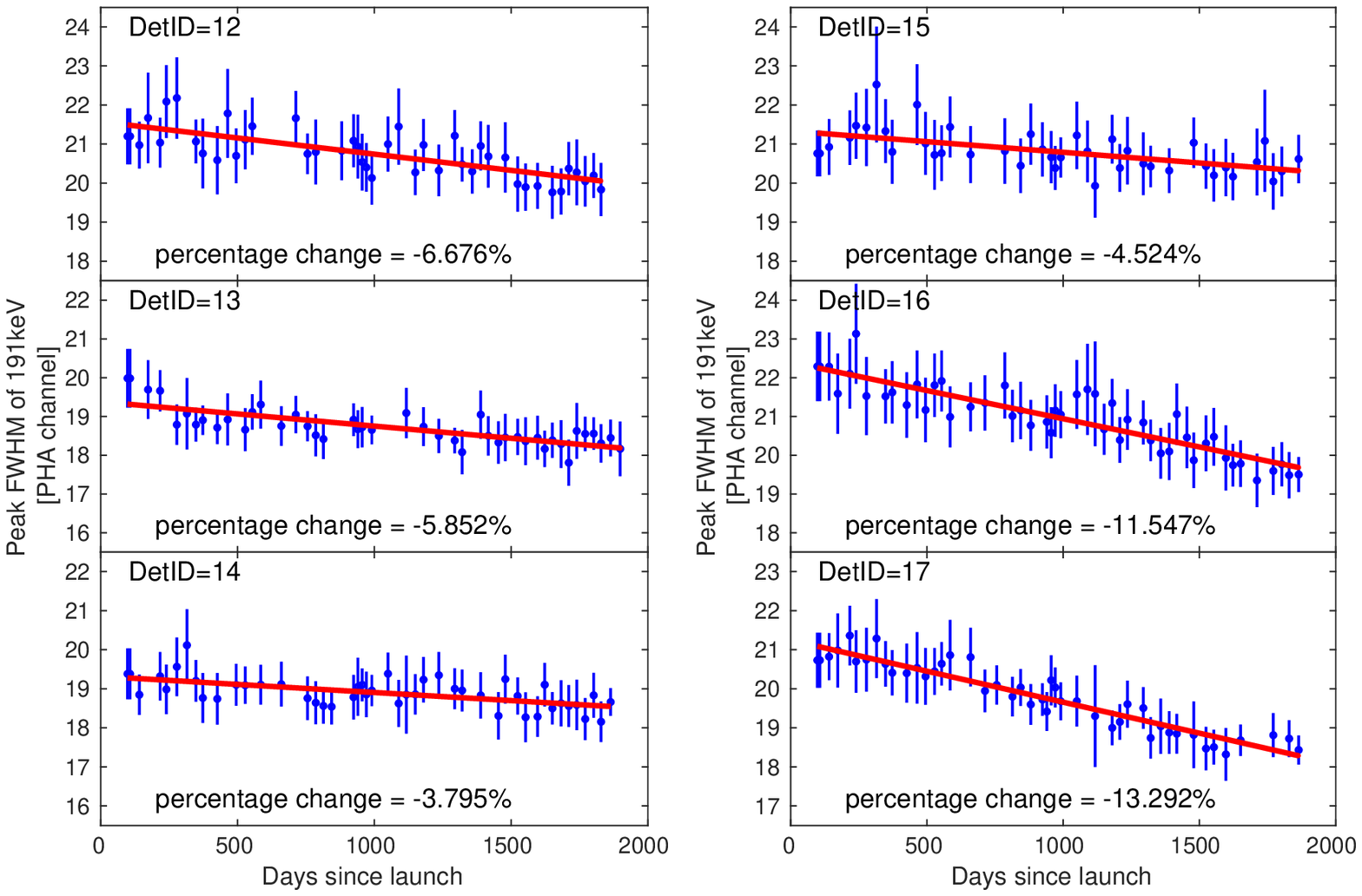}
\caption{The 191\,keV line resolution (in FWHM of PHA channel) for 18 HE/NaI(Tl) detectors as a function of time. Compared with the values just after launch, the percentage changes of the FWHM from Sep. 29, 2017 to Jun. 26, 2022 are shown in the subplots directly, and for most HE/NaI(Tl) detectors the energy resolutions are slightly getting better.}\label{fig6}
\end{figure}

In NOM, the gain of NaI(Tl) is automatically controlled by fixing the peak centroid of 59.5 keV calibration events to channel 50. Figure \ref{fig4_595peakinfo} shows the trend of the 59.5\,keV line peak centriod values and energy resolutions in full width at half maximum (FWHM) after launch. About 90 days after launch, the performance of HE/NaI(Tl) became stable. The gain variations were even less than 0.1$\%$ and the energy resolutions were getting better in the last five years. However, the parameters derived from the 59.5 keV events emitted by AGC sources can only represent a small local region of the corresponding NaI(Tl) detector. Here, we also use a background line around 191\,keV to monitor the gain variation of the overall NaI(Tl) detector with time. Figure \ref{fig5} and Figure \ref{fig6} show the evolution of the peak position and resolution in FWHM at 191 keV, respectively. The gain variations of all HE/NaI(Tl) detectors are less than 1$\%$ since launch. Also, for most of the HE/NaI(Tl) detectors the energy resolutions are slightly getting better. The results derived from the 191 keV line are highly consistent with that from 59.5 keV. 

The spectra of multi-pointing observations of the Crab nebula  were fitted by using the current energy response matrices of HE/NaI(Tl), which were calibrated in Oct. 2017, and if a residual larger than 2$\%$ in 28-120\,keV was observed, the energy response matrices would be re-calibrated. 

\subsubsection{CsI(Na) performance}\label{subsec3-2-2}
\begin{figure}[H]%
\centering
\includegraphics[width=1.0\textwidth]{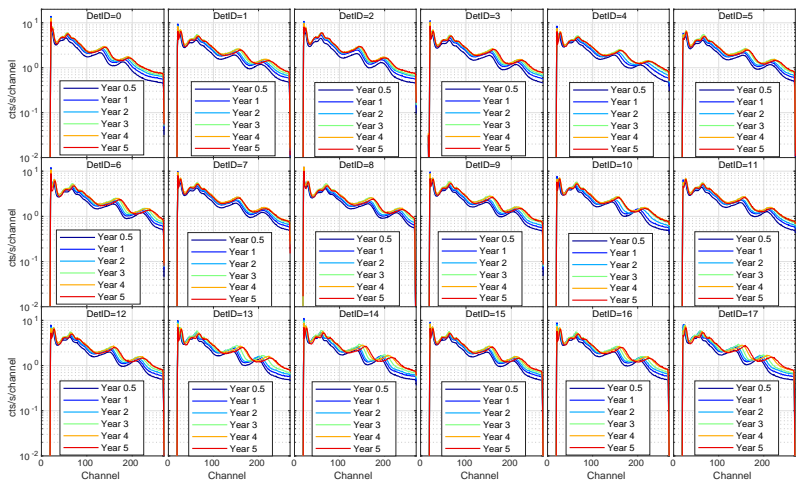}
\caption{Typical HE/CsI(Na) background spectra in NOM from blank-sky observations in different periods.}\label{fig7_CsIPHANOM}
\end{figure}

\begin{figure}[H]%
\centering
\includegraphics[width=1.0\textwidth]{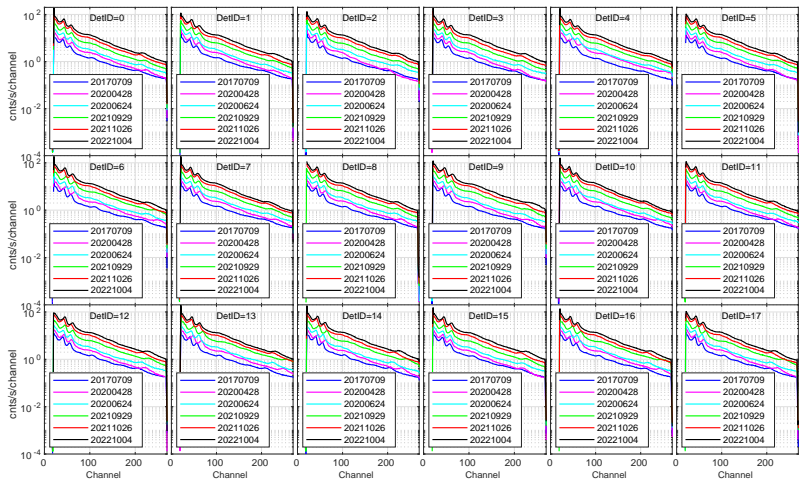}
\caption{Typical HE/CsI background spectra in LGM from earth occultation observations in different periods.}\label{fig8_CsIPHALGM}
\end{figure}

\begin{figure}[H]%
\centering
\includegraphics[width=0.32\textwidth]{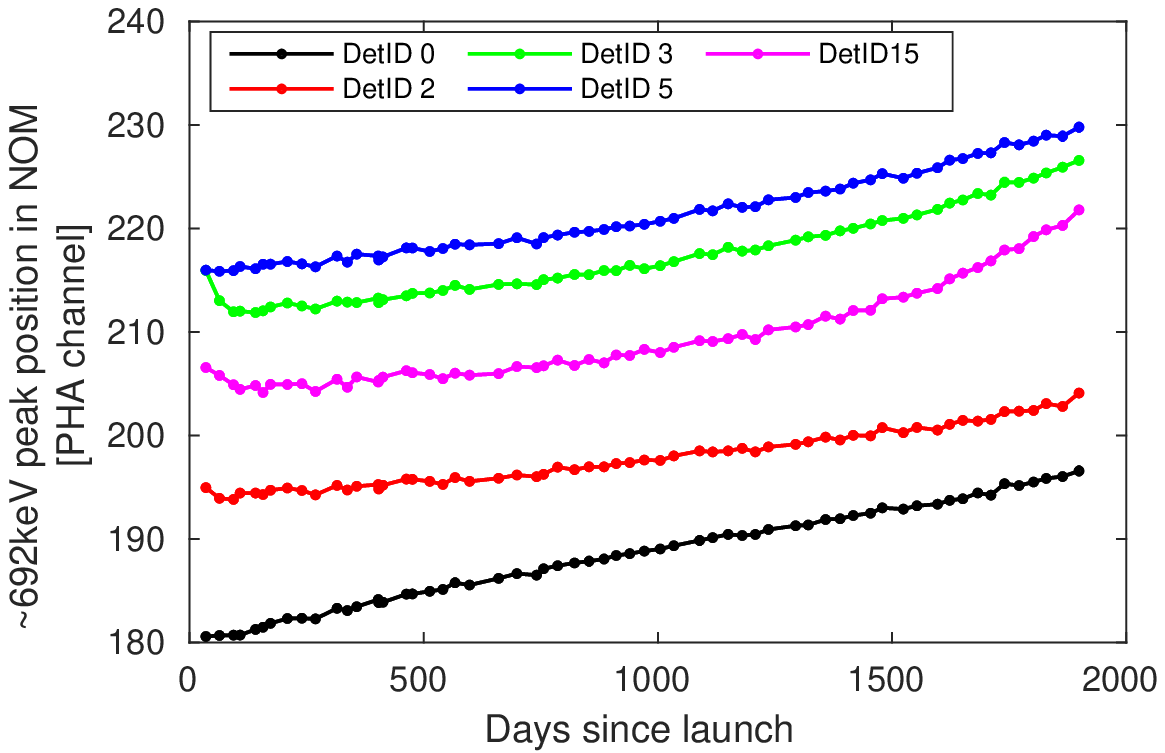}
\includegraphics[width=0.32\textwidth]{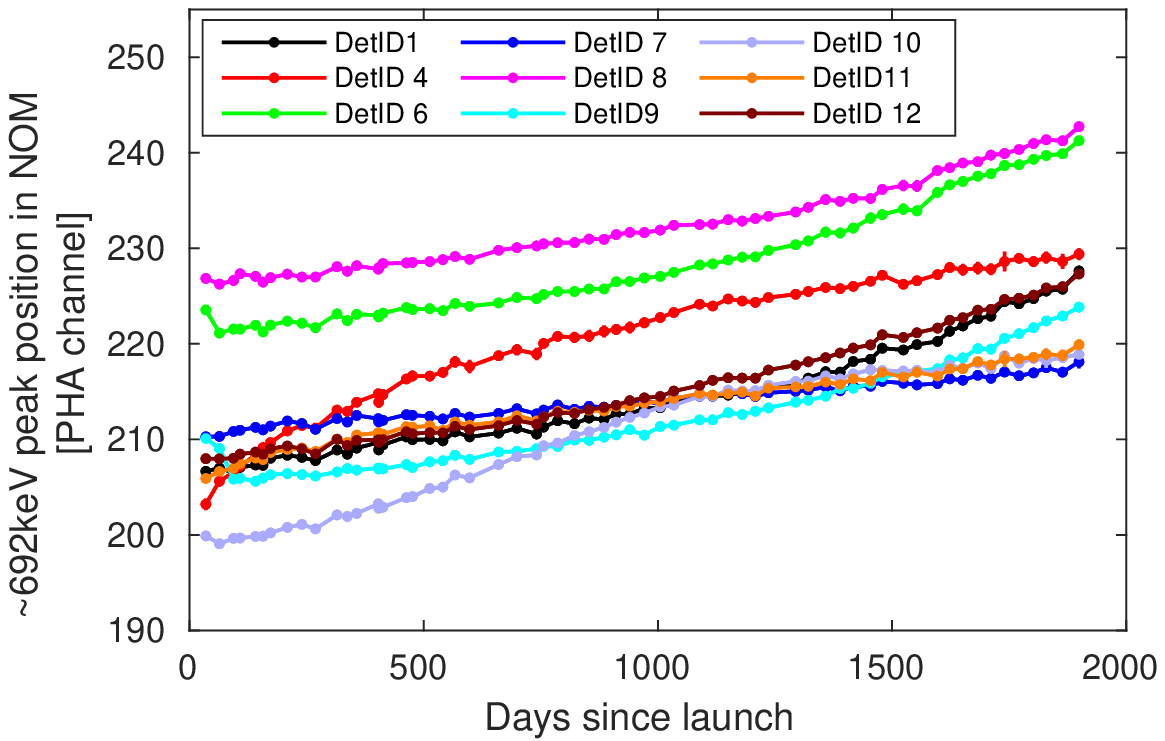}
\includegraphics[width=0.32\textwidth]{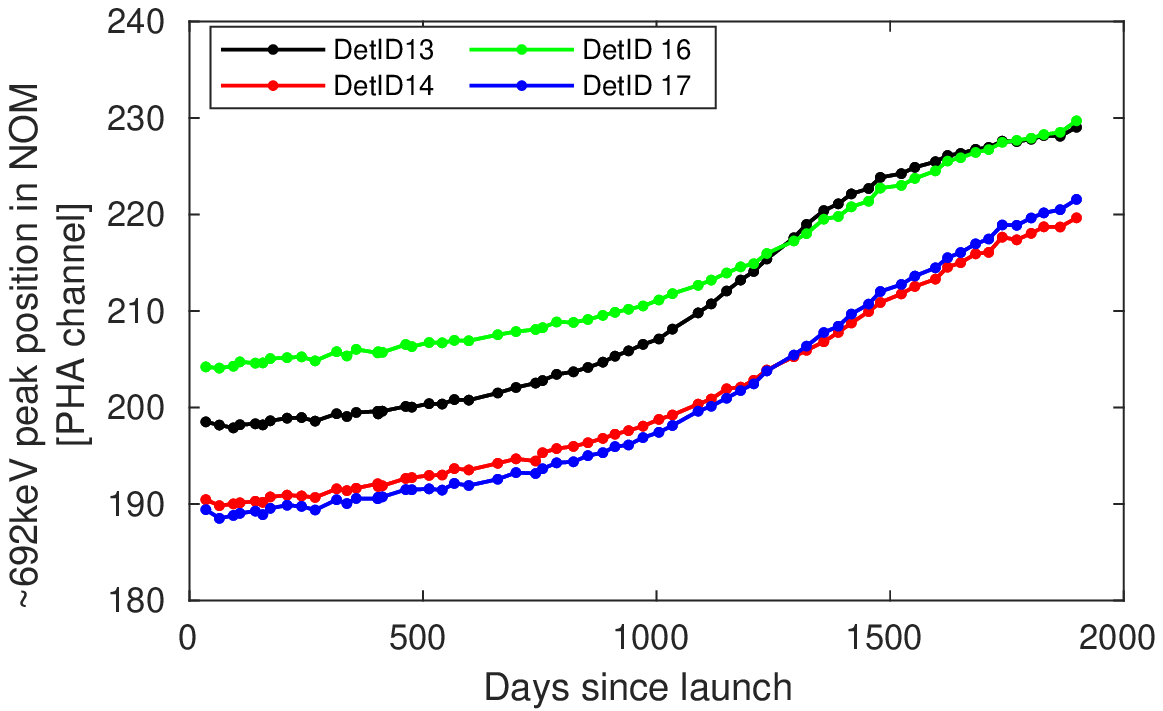}
\caption{The positions of $\thicksim$692\,keV peak centroid (in PHA channel) in NOM as a function of time.}\label{fig9_CsIpeakpositionNOM}
\end{figure}

\begin{figure}[H]%
\centering
\includegraphics[width=0.45\textwidth]{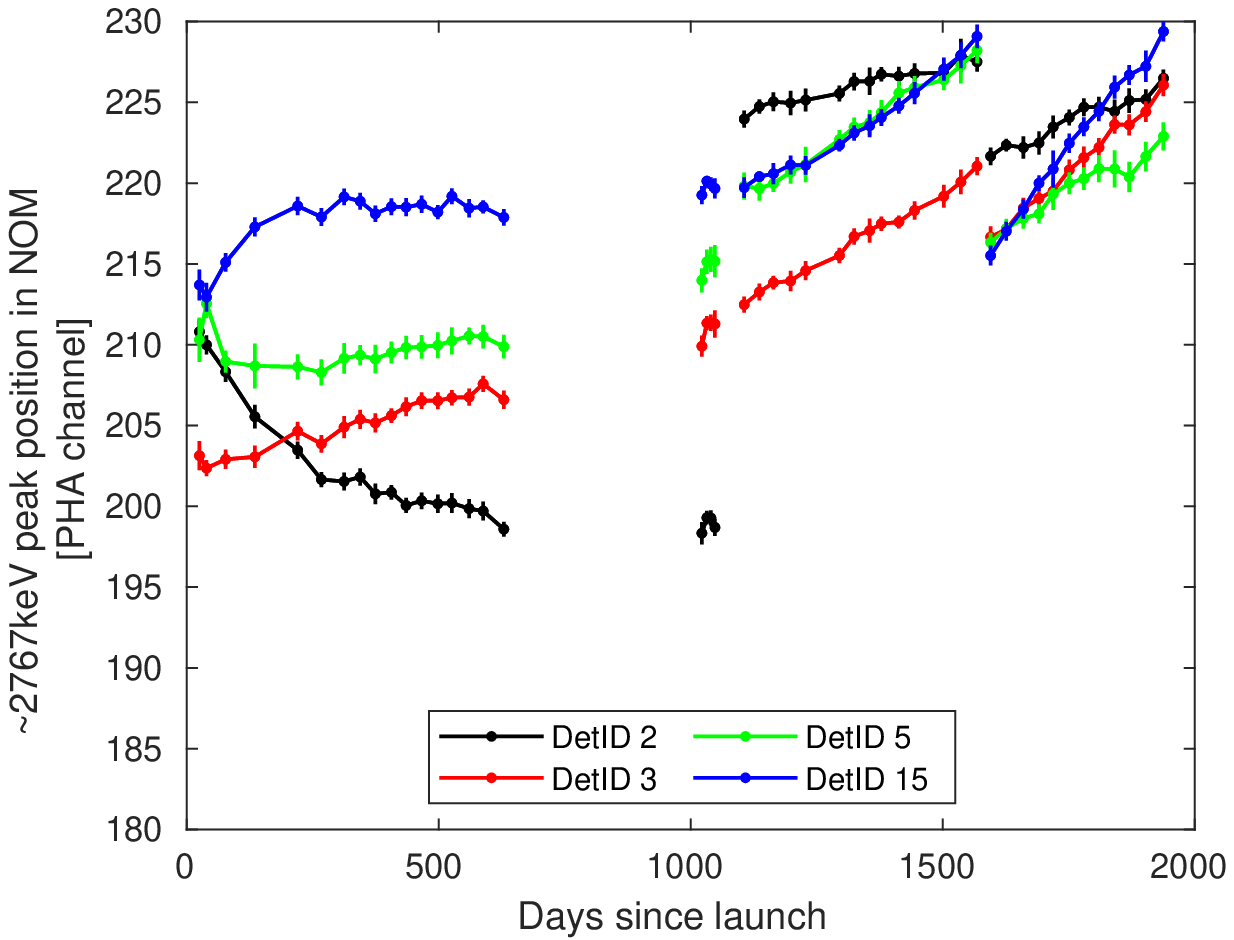}
\includegraphics[width=0.45\textwidth]{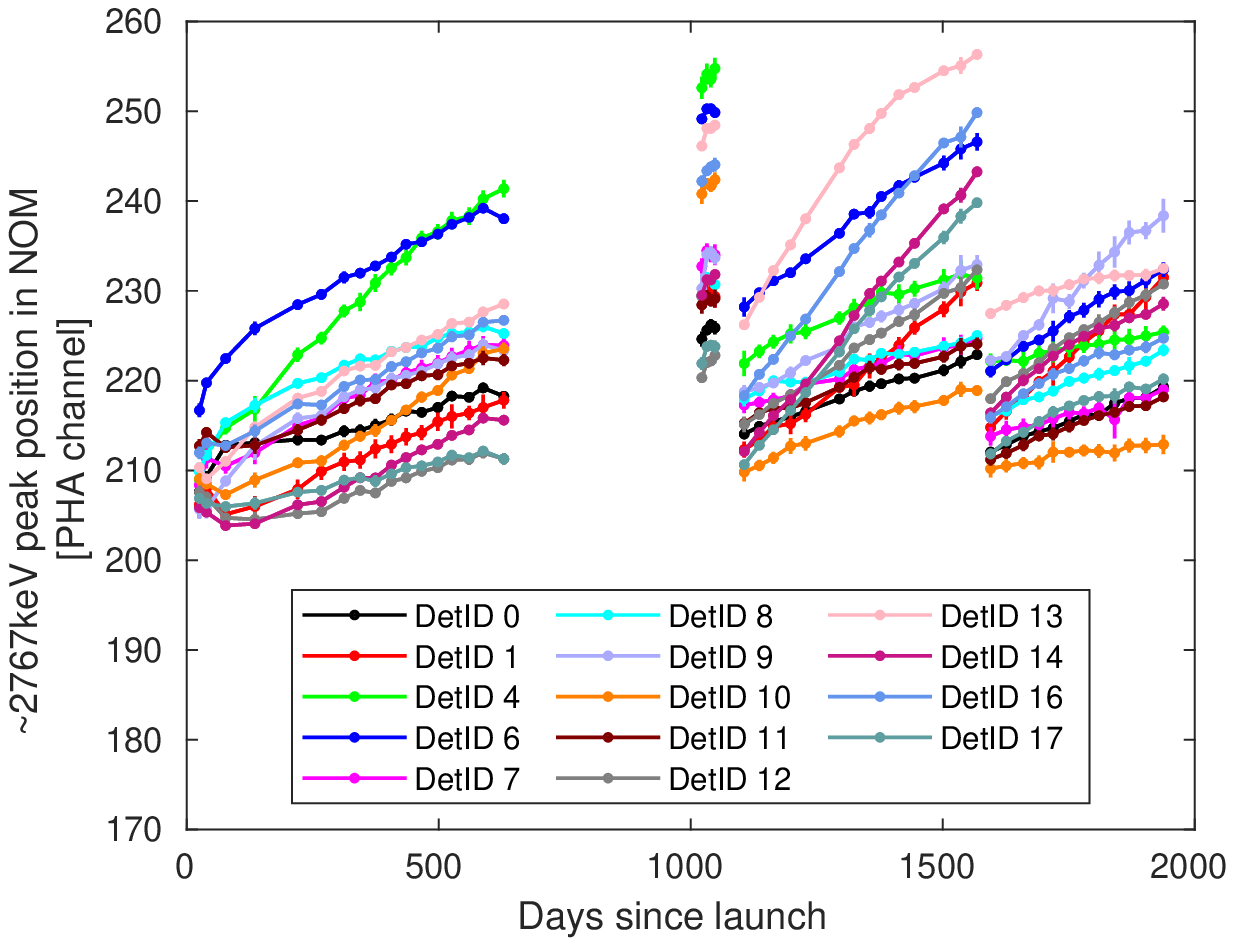}
\caption{The poisitions of $\thicksim$2767\,keV peak centroid (in PHA channel) in LGM as a function of time.}\label{fig10_CsIpeakpositionLGM}
\end{figure}

We can understand the evolution of the HE/CsI(Na) detector performances by monitoring their background PHA spectra during the last five years (see Figure \ref{fig7_CsIPHANOM} in NOM and Figure \ref{fig8_CsIPHALGM} in LGM). We can see that almost all the characteristic bumps of the spectra shift to the right over time, which indicates that the gain of the HE/CsI(Na) detector has increased continuously. For clarity, we plot the $\thicksim$692\,keV peak positions in NOM as a function of time in Figure \ref{fig9_CsIpeakpositionNOM} and the $\thicksim$2767\,keV peak positions in LGM as a function of time in Figure \ref{fig10_CsIpeakpositionLGM}. The $\thicksim$2767\,keV peak would shift out of the CsI(Na) energy range if the detector gain increased to a certain extent. So we had to adjust the PMT HVs in LGM in Jun. 2020 and Oct. 2021 respectively, to keep the energy range of CsI(Na) basically unchanged. That is why there are two obvious discontinuities in Figure \ref{fig10_CsIpeakpositionLGM}. From the left panel of Figure \ref{fig10_CsIpeakpositionLGM}, the gain of the detector DetID 2 decreased with time before Jun. 2020 but turned to increase after that.  The abnormal evolution behavior of DetID 2 is confusing and not understood well yet. Nevertheless, the data of DetID 2 can still be used normally in scientific research.

\begin{figure}[H]%
\centering
\includegraphics[width=1.0\textwidth]{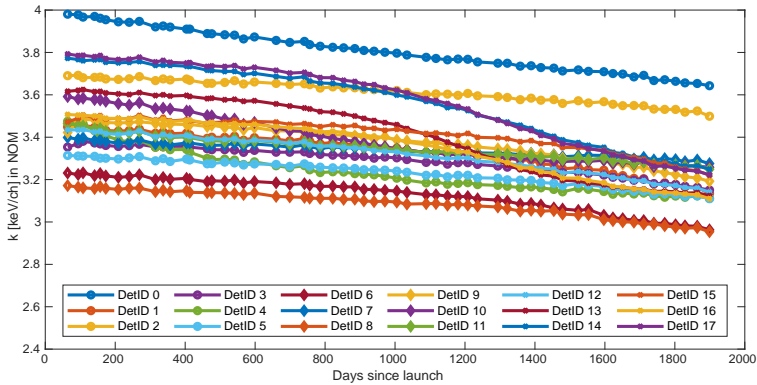}
\caption{Long-term monitor of the HE/CsI(Na) channel-energy relations in NOM.}\label{fig11_CsIslopeNOM}
\end{figure}

\begin{figure}[H]%
\centering
\includegraphics[width=1.0\textwidth]{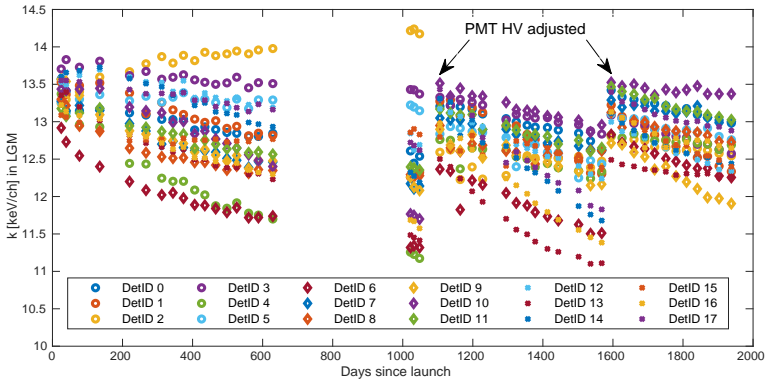}
\caption{Long-term monitor of the HE/CsI(Na) channel-energy relations in LGM.}\label{fig12_CsIslopeLGM}
\end{figure}

Because the gain of the HE/CsI(Na) detector changes significantly over time, we had to calibrate the channel-energy relation of the HE/CsI(Na) monthly. The energy responses of the HE/CsI(Na) detectors are calibrated using several background lines emitted by radioactive isotopes in the structure of the satellite, induced by  high energy particles (the energy of the lines are around 67,\,191,\,478,\,692,\,1182,\,1372,\,1729\,and\,2767\,keV)\cite{Luo2020,zhangjuan2020}. A linear function was adopted to describe the channel-energy relationship of the  HE/CsI(Na) in both normal mode and low-gain operation mode:
\begin{equation}
    E(x_{\rm c}) = k\cdot x_{\rm c} +b, \label{eq1}
\end{equation}
where $ E$ is the line energy in keV, $ x_{\rm c}$ is the line centroid position in channel, and $k$ represents the gain of each detector system in keV/ch. Figures \ref{fig11_CsIslopeNOM} and \ref{fig12_CsIslopeLGM} show the evolutions of the fitted slope index \textit{k} with time in the two operation modes, respectively. For most  CsI(Na) detectors, the coefficients of $k$ have been gradually decreasing in the first five years, except DetID 2 in LGM (see that in Figure \ref{fig12_CsIslopeLGM}).

\subsection{Performance evolution of PSD}\label{subsec3-3}
As described in section \ref{sec1}, HE employed NaI(Tl)/CsI(Na) phoswich detectors to observe compact objects in hard X-ray band of 20--250 keV. The PSD technology is essential to select NaI(Tl) events as the science data in NOM. The detailed technology was described and a figure of merit (FoM) was defined to characterize the discriminating capacity of a PSD used by HE in\cite{Liu2020}. The on-ground FoM values calibrated with a $^{133}$Ba radioactive source distributed between 2.6 and 3.0, showing a good discriminating capacity. 

\begin{figure}[H]%
\centering
\includegraphics[width=0.7\textwidth]{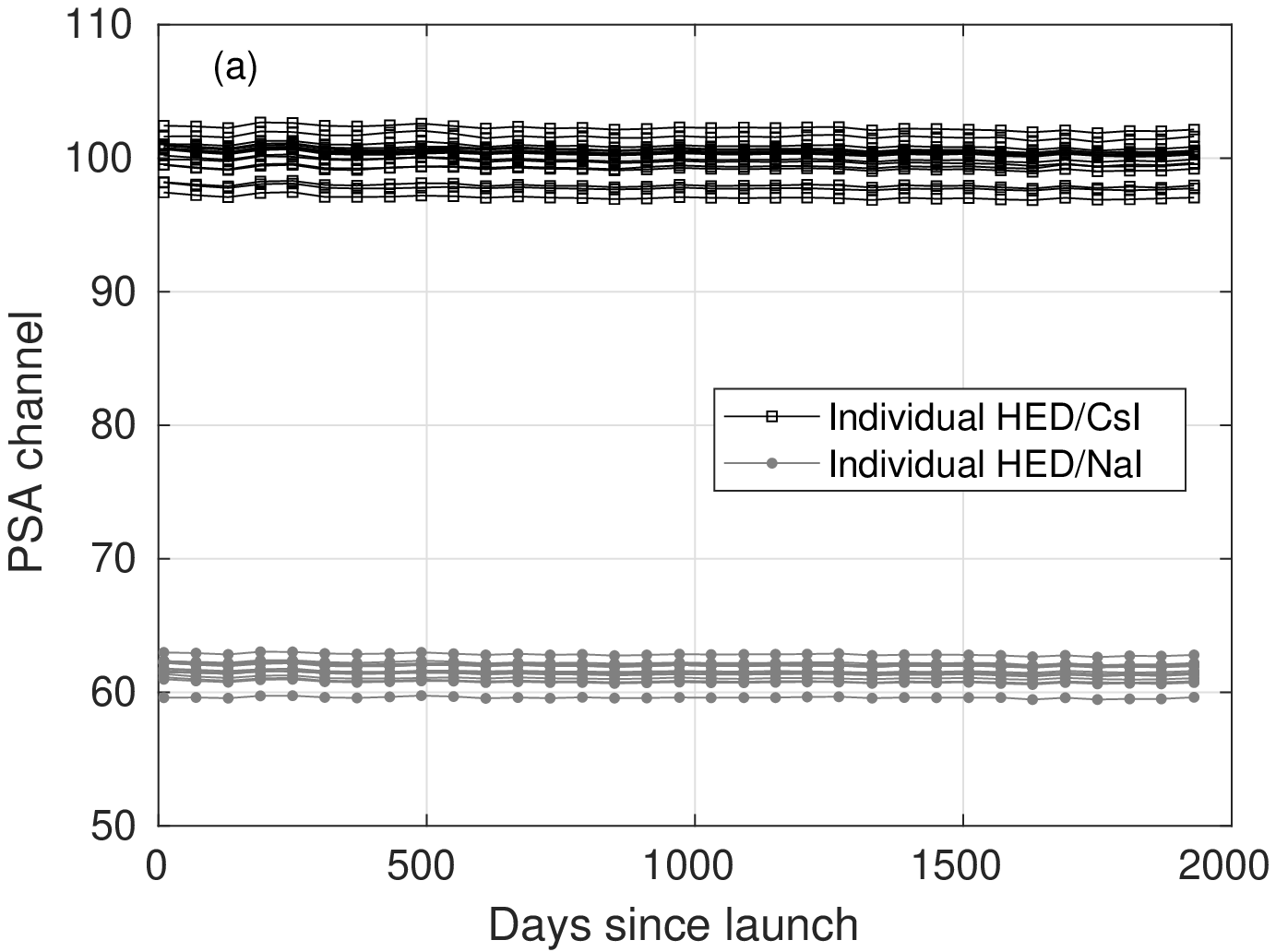}
\includegraphics[width=0.7\textwidth]{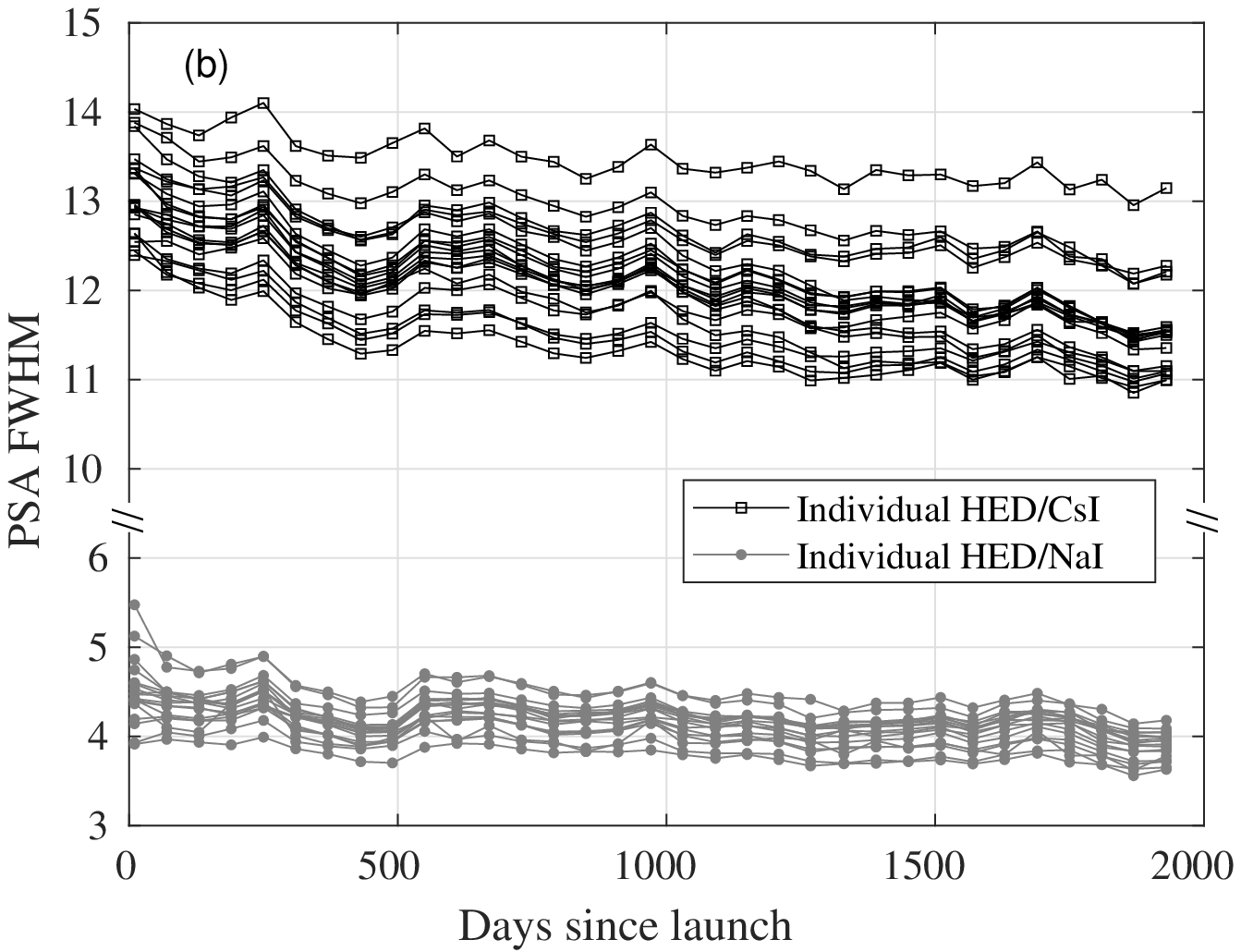}
\includegraphics[width=0.7\textwidth]{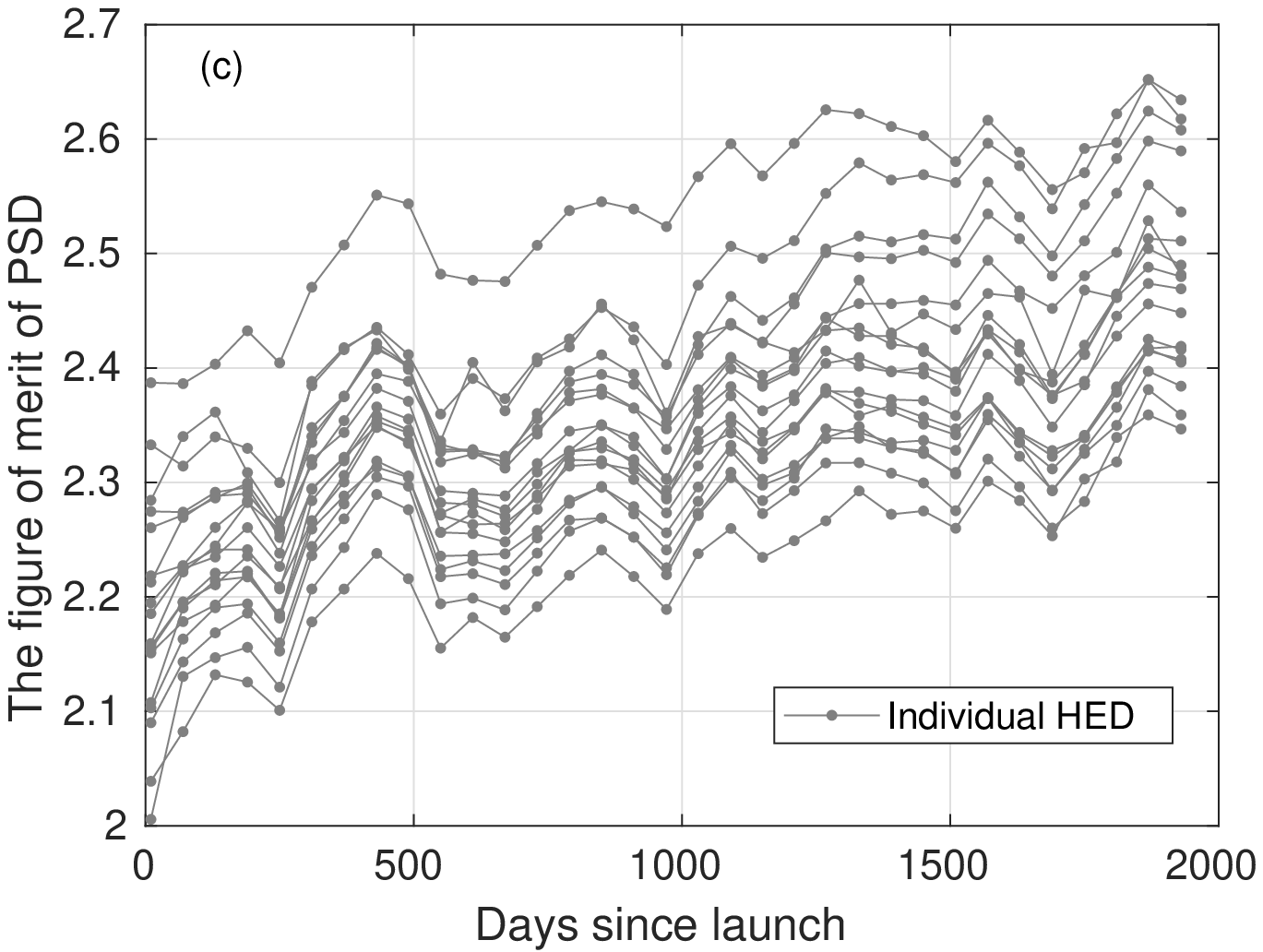}
\caption{Long-term monitor performances of PSDs. (a) the peak centroid in PSA channel evolution of NaI(Tl) (gray dots) and CsI(Na) (black squares) detectors over time. (b) the FWHM in PSA channel evolution of NaI(Tl) (gray dots) and CsI(Na) (black squares) detectors over time. (c) the FoM evolution of PSDs over time.}\label{fig13_PSD}
\end{figure}

To investigate the performance evolution of these PSDs, we select one day of background data of HE in earth occultation about every two months to make pulse width spectra. As described in\cite{Liu2020}, the pulse width spectrum consists of two obvious peaks corresponding to NaI and CsI events, respectively. We fit the two peaks with Gaussian functions to get the centroid values in PSA  channel and the peak widths in FWHM. Then we can get the FoM of each PSD with these parameters. Figure \ref{fig13_PSD} shows that the performances of the PSDs maintained high quality throughout the first five years. The peak centroids of eighteen NaI(Tl) detectors (gray dots) and CsI(Na) detectors (black squares) are fairly stable, as shown in Figure \ref{fig13_PSD}(a). In Figure \ref{fig13_PSD}(b), the peak widths of NaI(Tl) detectors (gray dots) show a decreasing trend of about 0.9$\%$ to 2.8$\%$ per year; the peak widths of CsI(Na) detectors (black squares) also show a decreasing trend of about 1$\%$ to 2.7$\%$ per year. According to the definition of FoM in\cite{Liu2020}, the stability of the two peaks' centroids and the decrease of the two peaks' widths will lead to the FoM of PSDs getting better. We can see that the FoMs are getting better at a level of about 1.2$\%$ to 2.8$\%$ per year in Figure \ref{fig13_PSD}(c). The in-orbit FoMs of PSDs are distributed between 2 and 2.7, which are slightly smaller than the values of on-ground calibration.

\begin{figure}[H]%
\centering
\includegraphics[width=0.7\textwidth]{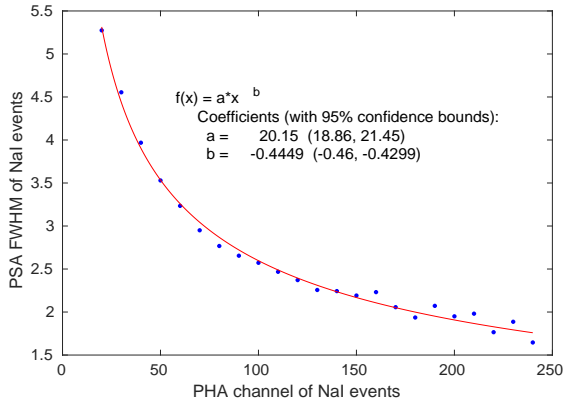}
\caption{The peak width of NaI events as a function of its pulse height, obtained from on-ground calibration on DetID 2 and all the HEDs have the similar behavior. The blue dots are the measured data and the red line is the fitting curve with a simple power function.  }\label{fig14_PSAwidthasPHA}
\end{figure}

Theoretically, the pulse widths of NaI(Tl) and CsI(Na) events should be independent of the energy of incident X-ray photons. But in the practical phoswich detector system, the pulse widths will show an obvious spread due to the fluctuations of fluorescence light output and also the noise of the electronics system. So the two peaks in the pulse width spectrum will be broadened into two Gaussian bumps rather than two lines. The peak widths must be taken into account when we analyse the FoM of a PSD. According to the on-ground calibration results, the peak widths of NaI events and CsI events decreased along with energy increase (see Figure \ref{fig14_PSAwidthasPHA}). The on-ground calibration source $^{133}$Ba has an emission line at about 80 keV, whereas the in-orbit background has a continuous energy spectrum covering 20--250 keV. So the FoM of a PSD is expected to become slightly worse with the in-orbit background data than that of on-ground calibration with $^{133}$Ba, as shown in Figure \ref{fig13_PSD}(c). Here we consider that the peak spread in the pulse width spectrum is mainly caused by the fluctuations of the fluorescence yielding and collection efficiency in the HED detectors. So the in-orbit FoMs of PSDs have been getting better as the gains of NaI(Tl) and CsI(Na) detectors have been  increasing in the first five years, as described in detail in section \ref{subsec3-2}.

\subsection{Performance evolution of dead-time}\label{subsec3-4}
\begin{figure}[H]%
\centering
\includegraphics[width=0.32\textwidth]{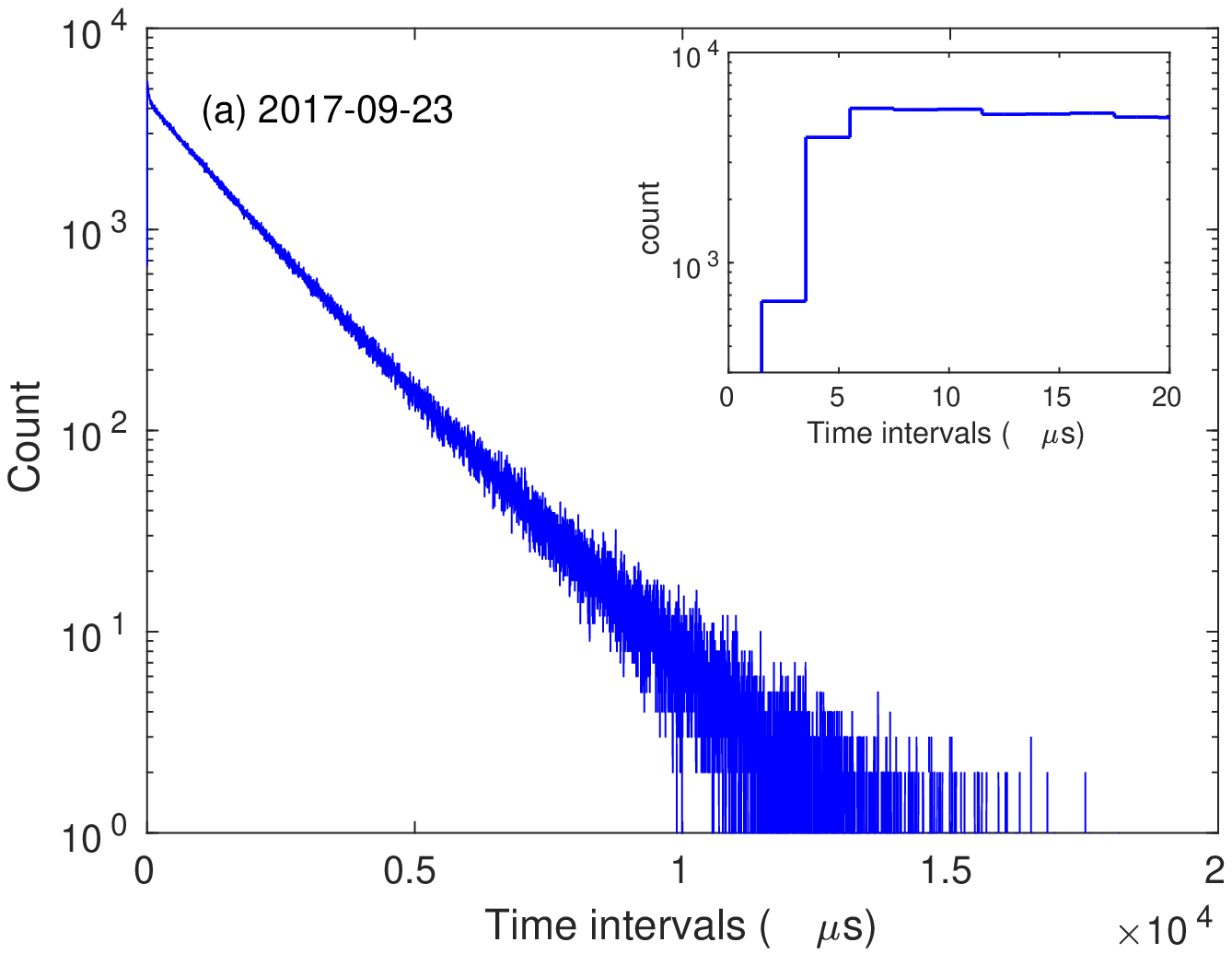}
\includegraphics[width=0.32\textwidth]{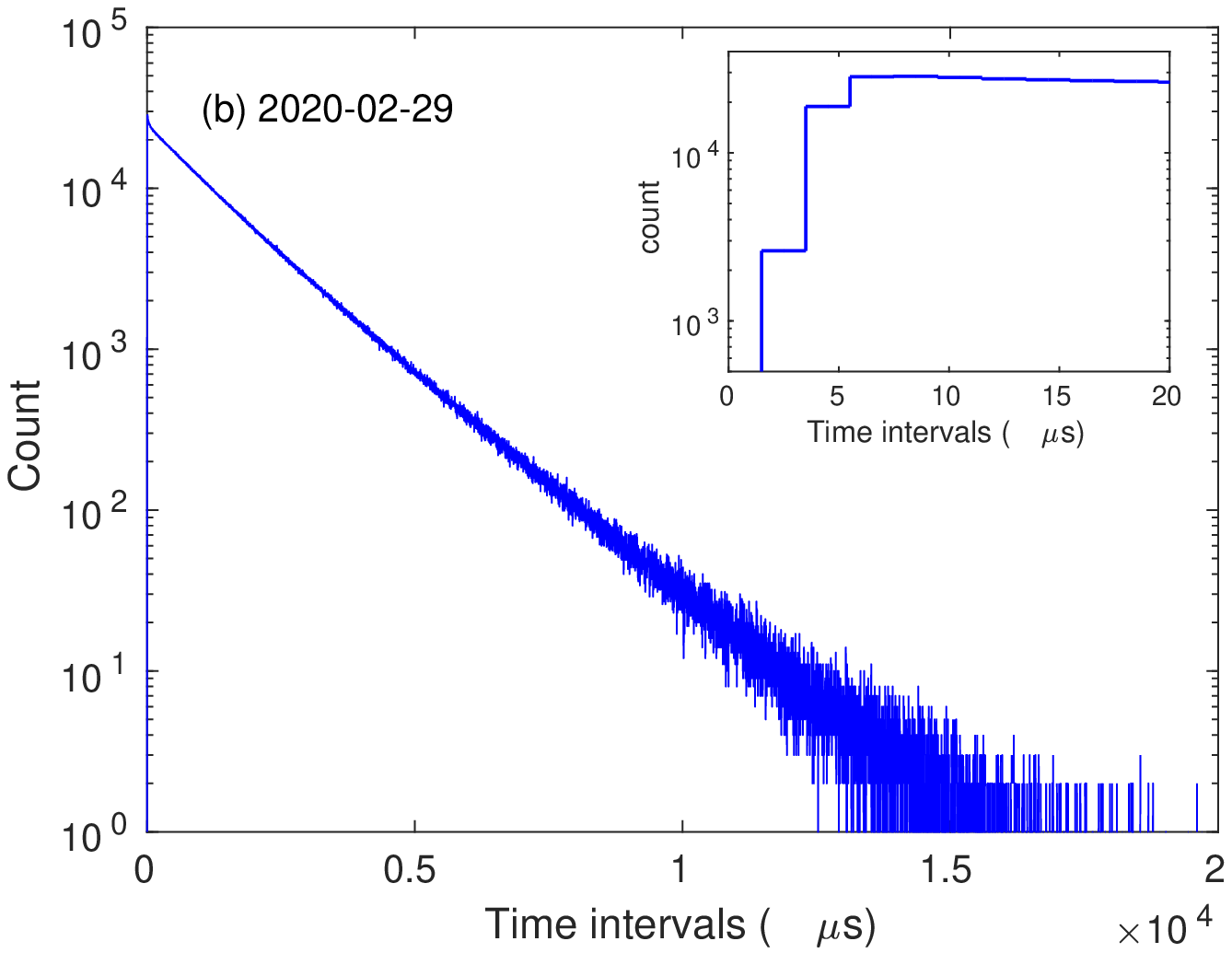}
\includegraphics[width=0.32\textwidth]{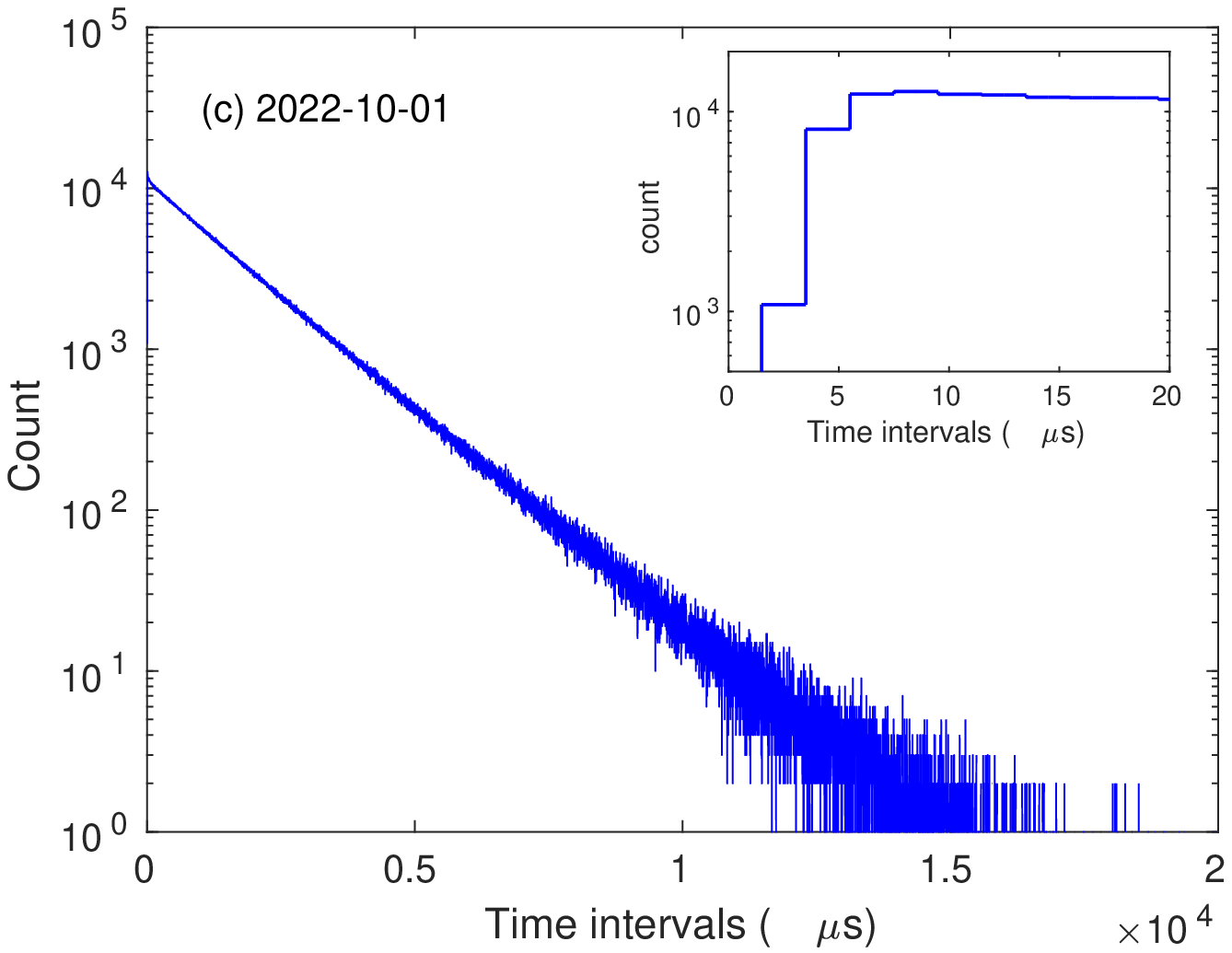}
\caption{Statistics of the arrival time interval between adjacent photons for HED with DetID=4 in different years. The insets show the details around the deviation point to clearly find out the maximum dead-time of one NaI event.}\label{fig15_evtTimeInterval}
\end{figure}

Dead-time is also an important parameter of the HED detector system. When HE observes bright sources (compact objects or transient bursts), the dead-time of the detector system must be corrected accurately to get the true light curves of the sources. The on-ground calibration results of dead-time were reported in detail\cite{Liu2020}. Xiao et al. (2020) proposed a parameterized model of the dead-time of the HE detector system to estimate the dead-time in any time scale. The method was successfully applied to bright transient source observations\cite{Xiao2020}. According to the results of on-ground calibration, the maximum dead-time of one NaI event is less than 8 $\mu s$\cite{Liu2020}. To investigate the evolution of dead-time, we select the background data of HE in earth occultation observed on Sept. 23, 2017, Feb. 29, 2020, and Oct. 1, 2022 as typical samples to represents the dead-time performance in the early, middle and late phase of the mission in the first 5 years. The spectra of the time interval of adjacent events follow negative exponential distribution perfectly, as shown in Figure \ref{fig15_evtTimeInterval}. The point deviating from the negative exponential distribution indicates the maximum dead-time of a good NaI event. From the inset in each panel of Figure \ref{fig15_evtTimeInterval}, we can find the dead-time of one NaI event kept unchanged at about 8 $\mu s$ during the first five years. It indicates that the electronics system of HE is very stable.

\begin{figure}[H]%
\centering
\includegraphics[width=1\textwidth]{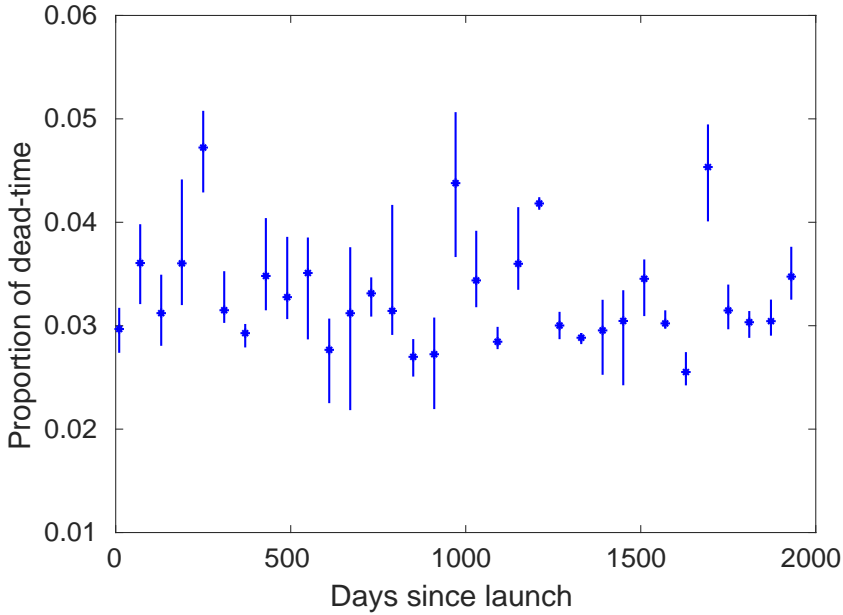}
\caption{Average proportion evolution of in-orbit dead-time with time during the first five years. The dots are the average proportion of dead-time in the specific days. Each error bar shows the maximum and minimum dead-time proportion in that day.}\label{fig16_deadtimePortion}
\end{figure}

We select the background data of HE in earth occultation about every two months. The data of each day are divided into several segments with every segment 2000 s long. We calculate the dead-time proportion of every segment. Then the evolution curve of the average proportion of in-orbit dead-time is plotted in Figure \ref{fig16_deadtimePortion}. The dots are the average proportion of dead-time on the specific days. Each error bar shows the maximum and minimum dead-time proportion in the day. It is easy to find that the accumulated proportion of dead-time is about 2 \% -- 6 $\%$, which is a fairly low value. The excellent dead-time performance makes HE capable of observing bright sources. Even so, some bright transient sources, especially the ultra-bright bursts, can make the HEDs saturated\cite{LiCK2021}, because the accumulated proportion of dead-time is indeed related to the in-orbit count rate of HED. 

\section{Summary}\label{sec4}
Three months after launch, the in-orbit performances of HE became stable. In NOM, the gain variations of all the HE/NaI(Tl) detectors are within 1$\%$ throughout the first 5 years, which should be credited to the AGC system. However, for the HE/CsI(Na) detectors in both NOM and LGM, the gain slowly changed with time and had to be calibrated monthly. The performances of PSDs have been maintained at a high level. The figures of merit of all PSDs are larger than 2 and show a slow increasing evolution. The performances of the dead-time do not show significant evolution in the first five years. In summary, HE has performed excellently and has obtained fruitful important science outcomes in the first five years\cite{LiTP2018, LiCK2021, Ma2021,Kong2022}. For now, HE is still in good health and well-calibrated status. We have been getting a comprehensive understanding of HE in the last five years. We therefore expect HE to achieve greater success in the era of multi-messenger astronomy, together with other instruments.

\section*{Declarations}
\begin{itemize}
\item This work used data from the \textit{Insight}-HXMT mission, a project funded by the China National Space Administration (CNSA) and the Chinese Academy of Sciences (CAS).  We gratefully acknowledge the support from the National Program on Key Research and Development Project (Grant No.2021YFA0718500) from the Minister of Science and Technology of China (MOST). The authors thank supports from the National Natural Science Foundation of China under Grants 12273043, U1838201, U1838202, U1938109, U1938102, U1938108. This work was partially supported by International Partnership Program of Chinese Academy of Sciences (Grant No.113111KYSB20190020))
\item \textbf{Conflict of interest}  On behalf of all authors, the corresponding author states
that there is no conflict of interest.
\end{itemize}

\bibliography{sn-bibliography}

\end{document}